\documentclass[a4paper,11pt,nohyper]{JHEP3}
\usepackage{graphicx}
\usepackage{multirow}

\def\d{\delta}

\def\t{\tau}

\def\S{\Sigma}

\def\del{\partial}              
   \let\d=\delta

 \let\t=\tau

\def\nn{\nonumber} \def\bd{\begin{document}} \def\ed{\end{document}}
\def\ds{\documentstyle} \let\fr=\frac \let\bl=\bigl \let\br=\bigr
\let\Br=\Bigr \let\Bl=\Bigl
\let\bm=\bibitem
\let\na=\nabla
\let\pa=\partial \let\ov=\overline
\newcommand{\be}{\begin{equation}}
\newcommand{\ee}{\end{equation}}
\def\ba{\begin{array}}
\def\ea{\end{array}}
\def\ft#1#2{{\textstyle{{\scriptstyle #1}\over {\scriptstyle #2}}}}
\def\fft#1#2{{#1 \over #2}}
\def\del{\partial}
\def\sst#1{{\scriptscriptstyle #1}}
 \def\oneone{\rlap 1\mkern4mu{\rm l}}
\def\ie{{\it i.e.\ }}
\def\via{{\it via}}
\def\semi{{\ltimes}}
\def\str{{\rm str}}
\def\Dm{{{D_{\sst{max}}}}}
\def\vac{ \left | 0 \right \rangle }
\def\kvac{ \left | k \right \rangle }

\def\sp{\; \; \;}

\def\bol{ \left | B (p^+) \right \rangle}
\def\bo1{ \left | B^0 (p^+) \right \rangle}

\def\bolt{ \left | B (p^+) \right \rangle_{\t}}

\def\boxl{ \left | B (x^-) \right \rangle}
\newcommand{\bea}{\begin{eqnarray}}
\newcommand{\eea}{\end{eqnarray}}

\def\<{ \langle }
\def\>{ \rangle }

\def\S{\Sigma}

\usepackage{amsmath,latexsym,amssymb,slashed}
\usepackage{graphicx}
\usepackage{psfrag}
\usepackage[latin1]{inputenc}
\usepackage{subfigure}

\renewcommand{\floatpagefraction}{0.6}
\renewcommand{\textfraction}{0.2}

\newcommand\ca{\mathcal{A}}
\newcommand\vp{\varphi}
\newcommand\beal{\begin{align}}
\newcommand\bbone{\ensuremath{\mathbbm{1}}}

\newcommand{\eq}[1]{\begin{equation}#1\end{equation}}
\newcommand{\spl}[1]{\begin{split}#1\end{split}}
\newcommand{\al}[1]{\begin{align}#1\end{align}}
\newcommand{\subeq}[1]{\begin{subequations}#1\end{subequations}}

\newcommand{\arXividhepth}[1]{\href{http://arxiv.org/abs/#1}arXiv:{\tt #1} [hep-th]}
\newcommand{\arXividother}[2]{\href{http://arxiv.org/abs/#1}arXiv:{\tt #1} [#2]}

\newcommand{\bg}[1]{\hat{#1}}
\newcommand{\wj}{\widetilde{J}}
\newcommand{\reo}{\mathrm{Re}~\!\omega}
\newcommand{\imo}{\mathrm{Im}~\!\omega}
\newcommand{\ads}{AdS_4}
\newcommand{\mcal}{\mathcal{M}}
\newcommand{\ccal}{\mathcal{C}}
\newcommand{\ncal}{\mathcal{N}}
\newcommand{\boxedeq}[1]{
\begin{equation}
\fbox{
\rule[0.7cm]{0pt}{0pt}
$#1$
\rule[-0.45cm]{0pt}{0pt}
}
\end{equation}
}

\def\d{\text{d}}

\def\slashchar#1{\setbox0=\hbox{$#1$}           
\dimen0=\wd0                                 
\setbox1=\hbox{/} \dimen1=\wd1               
\ifdim\dimen0>\dimen1                        
\rlap{\hbox to \dimen0{\hfil/\hfil}}      
#1                                        
\else                                        
\rlap{\hbox to \dimen1{\hfil$#1$\hfil}}   
/                                         
\fi}

\def\Re           {{\rm Re\hskip0.1em}}
\def\Im           {{\rm Im\hskip0.1em}}

\newcommand{\E}{\text{\tiny E}}

\newcommand{\tV}{{\widetilde{V}}}
\newcommand{\tH}{{\tilde{h}}}
\newcommand{\tm}{{{m}}}
\newcommand{\tmu}{{\tilde{\mu}}}
\newcommand{\trho}{{\tilde{\rho}}}
\newcommand{\tv}{{\tilde{v}}}
\newcommand{\calo}{\mbox{${\cal O}$}}
\newcommand{\cala}{\mbox{${\cal A}$}}
\newcommand{\dd}{\mathrm{d}}
\newcommand{\ra}{\rightarrow}
\newcommand{\calv}{\mbox{${\cal V}$}}
\newcommand{\calh}{\mbox{${\cal H}$}}
\newcommand{\calm}{\mbox{${\cal M}$}}
\newcommand{\abs}[1]{\left| #1 \right|}
\newcommand{\zetaa}{{\psi}}
\newcommand{\tr}{{\rm tr}\,}

\newcommand{\ky}[1]{{\color{blue}{#1}}}

\title{Generalized conformal structure, dilaton gravity and SYK}

\author{Marika Taylor  \\

Mathematical Sciences and STAG Research Centre, University of Southampton, \\
Highfield, Southampton, SO17 1BJ, UK.

\bigskip
 E-mail:
 \email{m.m.taylor@soton.ac.uk}}

\abstract{A theory admits generalized conformal structure if the only scale in the quantum theory is set by a dimensionful coupling. 
SYK is an example of a theory with generalized conformal structure and in this paper we investigate the consequences of this
structure for correlation functions and for the holographic realization of SYK. The Ward identities associated with the generalized conformal structure of SYK are 
implemented holographically in gravity/multiple scalar theories, which always have a parent AdS$_3$ origin. For questions involving only the graviton/running scalar sector, one 
can always describe the bulk running in terms of a single scalar but multiple running scalars are in general needed once one includes the 
bulk fields corresponding to all SYK operators. We then explore chaos in holographic theories with generalized conformal structure. The four point function explored by Maldacena, Shenker and Stanford exhibits exactly the same chaotic behaviour in any such theory as in holographic realizations of conformal theories i.e. the dimensionful coupling scale does not affect the chaotic exponential growth.} 

\newcommand{\bx}{\ensuremath{{\vec{x}}}}
\newcommand{\bk}{\ensuremath{{\vec{k}}}}
\newcommand{\bM}{\mathbf{M}}
\newcommand{\tps}[2]{\texorpdfstring{#1}{#2}}
\newcommand{\fpq}[2]{\ensuremath{{}_{#1} F_{#2}}}
\newcommand{\tfpq}[2]{\ensuremath{{}_{#1} \tilde{F}_{#2}}}
\newcommand{\tG}{\ensuremath{{\tilde{G}}}}

\begin{document}

\newcommand{\td}{\tilde}
 \newcommand{\bc}{\begin{center}}
 \newcommand{\ec}{\end{center}}
 \newcommand{\bfr}{\begin{flushright}}
 \newcommand{\efr}{\end{flushright}}
 \newcommand{\bfl}{\begin{flushleft}}
 \newcommand{\efl}{\end{flushleft}}
 \newcommand{\bt}{\begin{tabular}}
 \newcommand{\et}{\end{tabular}}

\section{Introduction and summary}

There has recently been considerable interest in the SYK \cite{Sachdev:1992fk,Kitaev} and related models - see for example \cite{Gross:2016kjj,Witten:2016iux,Fu:2016vas,Klebanov:2016xxf,Nishinaka:2016nxg,Peng:2016mxj,Krishnan:2016bvg,Turiaci:2017zwd,Li:2017hdt,Berkooz:2017efq,Krishnan:2017ztz,Peng:2017kro,Kanazawa:2017dpd}. The SYK model of one-dimensional fermions is a model that is solvable in the large $N$ limit and captures features of black holes including a large ground state entropy and a Lyapunov exponent saturating the chaos bound proposed in  \cite{Maldacena:2015waa}. 

The natural conjecture for a holographic dual to SYK is a background involving an asymptotically AdS$_2$ factor. As discussed in \cite{Maldacena:2016upp}, one would not expect the geometry to be precisely AdS$_2$, but ``nearly" AdS$_2$, with the bulk involving a running scalar. One of the aims of this work is to discuss this ``nearly" conformal invariance further, from the perspective of the underlying Ward identities defining both sides of the holographic correspondence.

Symmetries of additional compact directions in the bulk are associated with global symmetries of the field theory. The SYK model has no such global symmetries, suggesting that either the holographic dual is non-critical or that the compact directions have no isometries. Support for the former possibility follows from an analysis of the SYK spectrum, obtained by extracting the OPE from four point functions. The spectrum contains many low dimension operators \cite{Polchinski:2016xgd,Maldacena:2016hyu}, indicating that the bulk dual may involve low tension strings. Nevertheless one should be able to bootstrap a bulk description for SYK from knowledge of the SYK correlation functions, and it is this perspective that will be taken in this paper. 

The ``nearly" conformal two-dimensional gravity/dilaton theories are particular examples of holographic theories with generalized conformal structure. Generalized conformal structure
\cite{Jevicki:1998yr,Jevicki:1998qs,Jevicki:1998ub} is perhaps most easily defined in a field theory. A theory with conformal invariance has a traceless stress energy tensor up to conformal anomalies: $\langle T^i_i \rangle \sim 0$. In a theory with generalized conformal structure the scale invariance is broken at the quantum level only by a source for a scalar operator ${\cal O}_{\Phi}$:
\be
\langle T^{i}_i \rangle + (d - \Delta_{\Phi}) \Phi_s \langle {\cal O}_{\Phi} \rangle \sim 0, \label{ggc}
\ee
with $d$ the spacetime dimension and $\Delta_{\Phi}$ the dimension of the operator. It is therefore the (dimensionful) coupling $\Phi_s$ that drives the renormalization group flow; $\Phi_s$ is the only scale in the theory. (Later on in this paper we will generalize this concept to allow for a finite number of operators driving the flow.)

Generalized conformal structure underlies a number of holographic dualities, the most prominent of which are the D$p$-branes (with $p \neq 3$) and the fundamental string \cite{Itzhaki:1998dd}. In all such cases, in the regime where the supergravity description is valid the dynamics is controlled by the dimensionful running of the coupling. The holographic dictionary for such non-conformal brane backgrounds was developed in detail in \cite{Kanitscheider:2008kd} and will be reviewed in section \ref{four}. As we discuss later, a particular feature of the holographic dictionary is that dilaton/gravity models with generalized conformal structure can be understood in terms of dimensional reductions of AdS gravity on tori \cite{Kanitscheider:2009as}. These tori are not in general of integer dimension - the dimension is related to the running of the dilaton and the thermodynamic behaviour of the theory. 

In a theory with generalized conformal structure, operator correlation functions depend on the dimensionful coupling $\Phi_s$ as well as on the operator separation. For example, the Euclidean two point function for a generic scalar operator ${\cal O}$ of dimension $\Delta$ behaves as
\be
\langle {\cal O}(x) {\cal O} (0) \rangle = \frac{f(\Phi_s |x|^{d- \Delta_{\Phi}})}{|x|^{2 \Delta}} 
\ee
where $f$ is a function of the dimensionless coupling $\Phi_s |x|^{d- \Delta_{\Phi}}$. For a theory in which the driving operator is irrelevant, the correlator would admit an IR expansion
\be
\langle {\cal O}(x) {\cal O} (0) \rangle = \frac{1}{|x|^{2 \Delta}} \left ( f_0 +  f_1 \frac{\Phi_s}{|x|^{\Delta_{\Phi}- d}} + \cdots \right )  
\ee
i.e. a Laurent series in $\Phi_s/|x|^{\Delta_{\Phi} - d}$ with $(f_0,f_1,\cdots)$ dimensionless coefficients. In the deep IR limit $|x|^{\Delta_{\Phi} -d} \gg \Phi_s$ this two point function 
looks conformal; it is the existence of the series of corrections that distinguishes conformal behaviour from generalized conformal structure. Precisely this structure is found in the SYK two point function - see for example \cite{Kitaev,Maldacena:2016hyu,Jevicki:2016bwu}.

The generalized conformal structure underlying two dimensional dilaton/gravity models and the relation to AdS$_3$ gravity was discussed previously in \cite{Cvetic:2016eiv}, where the holographic dictionary between the bulk metric, gauge field and dilaton and the dual Hamiltonian ${\cal H}$, global current ${\cal J}$ and ${\cal O}_{\Phi}$ was derived. Further discussion of the relation between SYK and AdS$_3$ can be found in the recent work  \cite{Das:2017pif}.

In this paper we point out an implication of the holographic dictionary: the two point functions of these operators vanish (up to contact terms). This result follows directly from the Ward identities and can also be understood by reducing the two point function of the stress energy tensor of a 2d CFT on a circle. The Hamiltonian, global current and ${\cal O}_{\Phi}$ can however acquire (time independent) expectation values, as they indeed do in the thermal state. 
Both properties indicate that these particular operators cannot be used to probe chaotic behaviour in the theory via out of time correlators: according to \cite{Maldacena:2015waa} one needs to probe chaos using operators that have non-vanishing correlators and vanishing thermal expectation values. Holographically, this implies that one will need to include additional scalar fields with appropriate couplings to the running scalar to explore chaos. 

\bigskip

In the SYK model (before Gaussian averaging) the flow is not driven by a single operator but by multiple four fermion operators $\psi^i \psi^j \psi^k \psi^l$, each with a coupling $\lambda_{ijkl}$. This prompts the question of how to realise generalised conformal structure with multiple scalars holographically. In section \ref{five} we construct two dimensional models with multiple running scalars and show that these can again always be interpreted in terms of AdS gravity theories compactified on (non-integer) dimension tori. 

The uplifted dimension $(2 \sigma + 1)$  is determined by the thermodynamics of the theory e.g. the entropy scales with temperature as $T^{2 \sigma - 1}$. Hence (even for multiple dilatons) linear scaling of the entropy with temperature picks out a parent AdS$_3$ structure in the bulk realization of SYK. 

The SYK model after Gaussian averaging has equal sources ${\cal J}^2$ for bilocal operators i.e. the trace Ward identity is 
\be
{\cal H} = \frac{1}{4} {\cal J}^2 {\cal O} \qquad
\int dt {\cal O}(t) =  \int dt \sum_A {\cal O}_A(t)  \equiv \int dt_1 dt_2  \sum_{A} \tilde{\cal O}_A (t_1) \tilde{\cal O}_A(t_2)
\ee
where the operators $\tilde{\cal O}_A$ denote collectively $\psi_i \psi_j \psi_k \psi_l$ i.e. the sum over $A$ is equivalent to summing the 
indices $(i,j,k,l)$ over $N$.

Here the driving term can either be thought of as a single operator ${\cal O}$, dual to a single bulk scalar $\phi$, or as a set of operators 
${\cal O}_A$ (with identical sources) dual to a set of scalars $\phi_A$ . The required Ward identities can be realised holographically using 
\be
I = - N {\cal J} \int d^2 x \sqrt{g} e^{\sum_A \gamma_A \phi_A} \left ( R  + 2  \right ) = - N {\cal J} \int d^2 x \sqrt{g} e^{\phi} \left ( R  + 2  \right ).
\ee
As we show in section \ref{five}, both possibilities for the action result in a bulk geometry that admits an uplift to AdS$_3$. The two possibilities cannot be distinguished if one is interested in questions involving only the metric/scalar sector. If however one includes additional bulk fields, corresponding to other operators in the SYK theory, then one needs to take into account that these bulk fields will in general have different couplings to each of the scalars $\phi_A$. 

These couplings are determined by the OPEs of the operators $\tilde{\cal O}_A$ with the operator under consideration. We argue in section \ref{seven} that the requirement of preserving the generalized conformal structure constrains the interactions between generic bulk fields and the metric/running scalars in the holographic action. These constraints are such that there is always a parent AdS$_3$ description of the theory.

\bigskip

Next we proceed to discuss the characterisation of chaos via out of time four point functions in a holographic realisation of a non-conformal theory with generalized conformal structure.  We consider holographic non-conformal theories in general dimensions, thus including not just putative duals to SYK but also non-conformal Dp-branes, fundamental strings etc. 

The bulk action is then of the generic form \cite{Kanitscheider:2008kd}
\be
I = - N^a L^b \int d^{d+1} x \sqrt{g} e^{\gamma \phi} \left ( R + \beta (\partial \phi)^2 + C \right ) \label{non-conformal}
\ee
where $(\gamma,\beta,C)$ are related in such a way that the equations admit AdS$_{d+1}$ solutions with the scalar running as $\alpha \log (\rho)$ and $(a,b)$ are constants. This action can be uplifted to an AdS action in $(2 \sigma + 1)$ dimensions where $2 \sigma = d -  \alpha \gamma$:
\be
I = - N^a \int d^{2 \sigma + 1} x \sqrt{G} \left ( R(G) + 2 \sigma (2 \sigma + 1) \right ).
\ee
Here $L^b$ is interpreted in terms of the volume of the $(2 \sigma - d)$-dimensional torus on which the theory is compactified and $L$ can be interpreted as the length scale set by the dimensionful coupling i.e. the length scale associated with the coupling $\Phi_s$ appearing in \eqref{ggc}. 

The parameter $N$ is related to the number of degrees of freedom in the dual theory. In top down realizations, all terms originating from supergravity will have the same $N$ scaling, with loop and $\alpha'$ corrections subleading in $N$. The constants $(a,b)$ relevant for non-conformal Dp-branes and fundamental strings can be found in \cite{Kanitscheider:2008kd}. For the conformal cases ($\beta = \gamma =0$) of M5-branes, D3-branes and M2-branes, the constant $a$ is $3$, $2$ and $3/2$ respectively. 

As we discussed above, probing chaos requires scalar operators that do not acquire thermal expectation values i.e. we need to include additional fields to the graviton and running scalars. Let $V$ and $W$ be such generic scalar operators that do not acquire thermal expectation values, with corresponding bulk scalar realisations $\varphi_V$ and $\varphi_W$ respectively. We can then characterize chaos in terms of the normalized out of time four point function 
\be
f(t) = \frac{ {\rm Tr} \left ( y V(0) y W(t) y V(0) y W(t) \right )}
{ {\rm Tr} \left ( y^2 V(0) y^2 V(0) \right ) {\rm Tr} \left (y^2 W(t) y^2 W(t) \right )} \label{4pf-norm}
\ee
where $y$ moves an operator a quarter of the way around the thermal circle. 

In a holographic realization of a conformal theory, it was argued in \cite{Maldacena:2015waa} that 
\be
f(t) = 1 - c \; G \exp \left ( \frac{2 \pi t}{\beta} \right ) + \cdots \label{con-result}
\ee
where $c$ is a dimensionless order one constant, $G$ is the (dimensionless) Newton constant and $\beta$ is the inverse temperature. The ellipses denote subleading contributions to the four point function. The first term is associated with the disconnected contribution to the four point function whereas the second term is associated with connected contributions. The latter are dominated by exchanges in which particles pass very close to the black hole horizon and so have large boosted energy \cite{Shenker:2013pqa,Roberts:2014isa,Roberts:2014ifa}: if the energy of the exchanged particle is of order one asymptotically, the energy is boosted by a factor of $\exp \left ( \frac{2 \pi t}{\beta} \right )$ close to the horizon. This boosted energy is comparable to the black hole energy when $\exp \left ( \frac{2 \pi t}{\beta} \right ) \sim 1/G$.  

The Newton constant $G$ scales with $N$ as $1/N^a$; note that $a$ is not in general two as was stated in \cite{Maldacena:2015waa}. In fact, a straightforward way of understanding the $N$ scaling
of the terms in \eqref{con-result} is via the $N$ scaling of the correlators. In a large $N$ CFT we usually normalise two point functions to one, with all higher point functions scaling as negative powers of $N$. In the holographic (supergravity) normalization all connected correlation functions are normalized as $N^a$. Therefore the disconnected contribution (which behaves as the square of a two point function) behaves as $N^{2a}$, and the ratio of connected to disconnected contributions goes as $1/N^a$. 

In a holographic theory with generalized conformal structure, one might expect that the normalized four-point function \eqref{4pf-norm} would depend on the dimensionful scale $L$ as well as $N$. However, we argue in this paper that the result \eqref{con-result} still holds, although subleading terms in \eqref{con-result} would depend explicitly on $L$. 
There are several ways to show this. Firstly, we explained above that such holographic theories can always be uplifted to AdS gravity theories, for which the arguments of \cite{Maldacena:2015waa} apply. Note that the dimension of the uplifted spacetime appears in \eqref{con-result} only indirectly through the temperature. 

Another way to understand why \eqref{con-result} applies to holographic theories with action \eqref{non-conformal} is via scattering in the $d$-dimensional black hole geometry. 
In the dual frame the black hole geometries are asymptotically AdS and so again the energy is boosted by a factor of $\exp \left ( \frac{2 \pi t}{\beta} \right )$ close to the horizon. The holographic dictionary implies that the energy of the exchanged particle is of order $L^b$ asymptotically while the energy of the black hole itself goes as $L^b N^a$. Therefore the boosted energy is comparable to the black hole energy when $\exp \left ( \frac{2 \pi t}{\beta} \right ) \sim N^a$ i.e. the dimensionful factors cancel.

\bigskip

The plan of this paper is as follows. In section \ref{two} we briefly review relevant features of the SYK model. In section \ref{three} we define generalized conformal structure and explain how this structure underlies non-conformal brane theories (maximally supersymmetric Yang-Mills) and SYK. We discuss the holographic realization of generalized conformal structure in section \ref{four} and show that the operators dual to 2d dilaton gravity have trivial correlation functions. In section \ref{five} we discuss 2d dilaton gravity models with multiple scalars and show that any such model admitting generalized conformal structure has a parent AdS theory. We show how such models arise from reductions of conformal field theories in section \ref{six}, using the example of ${\cal N} = 4$ SYM reduced on a torus. In section \ref{seven} we discuss interaction terms and chaos in holographic realizations of theories with generalized conformal structure and in section \ref{eight} we conclude.

\section{Review of relevant features of SYK model} \label{two}

The SYK model  \cite{Sachdev:1992fk,Kitaev} consists of $N$ fermions $\psi_i$ in one dimension, with Lagrangian given by
\be
{\cal L} = \sum_{i} \psi_i \partial_{t} \psi_i + \sum_{ijkl} \lambda_{ijkl} \psi_i \psi_j \psi_k \psi_l.
\ee
Here the couplings $\lambda_{ijkl}$ are drawn randomly from a Gaussian ensemble and have zero average, with the width of the Gaussian
being ${\cal J}/ N^{3/2}$. The parameter ${\cal J}$ sets the scale for the theory and has dimension one, since the fermions are dimension zero. 
Generalisations to interactions involving $q$ interactions (with $q$ even) are discussed in \cite{Maldacena:2016hyu}; in what follows we will be interested in the case of $q > 2$. The $q=2$ 
case was also discussed in \cite{Maldacena:2016hyu} but the behaviour is qualitatively different in this case. 
More details about the properties of SYK can be found in \cite{Sachdev:1992fk,Kitaev,Maldacena:2016hyu,Jevicki:2016bwu,Jevicki:2016ito,Garcia-Garcia:2016mno,Garcia-Garcia:2017pzl,Gross:2017hcz}.

The Hamiltonian ${\cal H}$ can be expressed as 
\be
{\cal H} = \sum_{ijkl} \lambda_{ijkl} \psi_i \psi_j \psi_k \psi_l \equiv \sum_{ijkl} \lambda_{ijkl} {\cal O}_{ijkl}
\ee
where for future use we denote by ${\cal O}_{ijkl}$ the four fermion (composite) operators. The Hamiltonian satisfies
\be
\partial_{t} {\cal H} = \sum_{ijkl} ( \partial_{t} \lambda_{ijkl} ) {\cal O}_{ijkl}
\ee
i.e. the Hamiltonian is conserved when the couplings are time independent. 

In the large $N$ limit the theory is solvable with the bilocal field
\be
G(\tau_1, \tau_2) = \frac{1}{N} \sum_{i} \psi_i (\tau_1) \psi_i (\tau_2) \label{bilocal}
\ee
becoming classical in this limit. (Note that throughout this paper we use $t$ to denote Lorentzian time and $\tau$ to denote imaginary time.)
For large values of ${\cal J} (\tau_1 - \tau_2) \equiv {\cal J} \tau$
\be
G(\tau_1,\tau_2) \propto \tau^{-2 \Delta}
\ee 
where $\Delta  = 1/q$ and $q$ is (as above) the number of fermions in each interaction. We will discuss this behaviour further in the next section. 

Using the reparameterization
\be
\tau \rightarrow f(\tau) = \tan \left ( \frac{\tau \pi}{\beta} \right )
\ee
we can also obtain the finite temperature two point function 
\be
G(\tau_1,\tau_2) \propto \left ( \frac{\pi}{\beta \sin \left ( \frac{ \pi \tau}{\beta} \right )} \right )^{2 \Delta}
\ee
with $\beta$ the inverse temperature. 

The SYK model has the feature that the entropy is of order $N$ at zero temperature. (It is this feature that first suggested a connection with AdS$_2$ black holes in string theory \cite{Sachdev:2010um,Sachdev:2015efa}). 
More precisely, the partition function behaves as 
\be
\log Z = - \beta E_0 + S_0 + \frac{c}{2 \beta}
\ee
where the ground state energy $E_0$ and the zero temperature entropy $S_0$ scale with $N$. The specific heat $c$ can be shown to be proportional 
to $N/{\cal J}$ by analysing reparameterizations. 

Chaotic behaviour may be explored by computing (appropriately time ordered) four point functions 
\be
\sum_{ij} \langle  \psi_i (\tau_1) \psi_i (\tau_2) \psi^j (\tau_3) \psi^j (\tau_4)
\ee
averaged over $i,j$. As a $1/N$ expansion such four point functions have the form 
\be
G(\tau_{12}) G(\tau_{34}) + \frac{1}{N} F (\tau_1,\tau_2,\tau_3,\tau_4) + \cdots
\ee
i.e. the leading in $N$ contribution is the disconnected contribution. The first subleading contribution can be expressed in terms of a function ${\cal F}(\chi)$ of a single cross ratio $\chi$, see \cite{Maldacena:2016hyu}. 

The out of time correlator used to explore chaos is 
\be
{\rm Tr} \left ( y \psi^i(t) \psi^i (0) y \psi^i(t) y \psi^j (0) \right )
\ee
where the operator $y$ separates the fermions by a quarter of the thermal circle \cite{Maldacena:2015waa}. (Here we again use $t$ to denote real time, with $\tau$ reserved for Euclidean time.) The $1/N$ contribution to this correlator grows in time with a Lyapunov exponent
\be
\lambda_L = \frac{2 \pi}{\beta},
\ee
in agreement with black holes, thus saturating the proposed chaos bound. 

In addition to the contributions growing exponentially with time, there are finite (connected) contributions to the four point function that contain information about composite operators in the OPE of $G(\tau)$. These composite operators form a tower of states with approximately integer spacing, suggesting that any holographic dual would need a large number of light states; see \cite{Polchinski:2016xgd,Maldacena:2016hyu} for further discussion of the spectrum. 

\section{Conformal symmetry versus generalized conformal structure} \label{three}

Consider a conformally invariant quantum field theory. A scalar operator ${\cal O}$ of scaling dimension $\Delta$ in a CFT has a two point function
\be
\langle {\cal O} (x) {\cal O} (0) \rangle = \frac{f(g_i,N,\cdots)}{|x|^{2 \Delta}}
\ee
where $f$ is a function of the dimensionless couplings of the theory $g_i$ and of other dimensionless parameters such as the rank $N$ of a gauge group. In practice it is usually convenient to normalise the two point function such that $f = 1$ over the entire moduli space of the CFT. For later use, it is useful to note that this is not the natural holographic normalization: the holographic normalization is proportional to the supergravity action normalization. For example, for $AdS_5 \times S^5$ the two point functions are normalized as $N^2$ and the interaction terms in supergravity give the leading $N$ terms in higher point functions as order $N^2$ also. 

Now let us reverse the logic: suppose that a two point function of a scalar operator behaves as 
\be
\langle {\cal O} (x) {\cal O} (0) \rangle \sim \frac{1}{|x|^{2 \Delta}}
\ee
in a specific limit e.g. large separations $x$. This does not in itself imply that the theory is scale invariant in this limit, even if all scalar operator two point functions exhibit scaling behaviour. 

The particular context that we have in mind in this work is the presence of underlying generalized conformal structure, introduced in \cite{Jevicki:1998yr,Jevicki:1998qs,Jevicki:1998ub} in the context of non-conformal D-branes. 
Let us suppose that conformal invariance is broken by a source $\Phi_s$ for a scalar operator ${\cal O}_{\Phi}$ of dimension $\Delta_{\Phi}$, and that this structure is respected at the quantum level (up to possible quantum anomalies, analogous to the Weyl anomaly of a conformal field theory). Then the dilatation Ward identity for the theory is
\be
\langle T^i_i \rangle + (d- \Delta_{\Phi}) \Phi_s \langle {\cal O}_{\Phi} \rangle = {\cal A}
\ee
where $T_{ij}$ is the stress energy tensor, $d$ is the spacetime dimension and ${\cal A}$ is the anomaly. Similarly the diffeomorphism Ward identity is 
\be
\nabla^i \langle T_{ij} \rangle + \partial_j \Phi_s \langle {\cal O}_{\Phi} \rangle = 0, 
\ee
where $\nabla_i$ is the covariant derivative and we assume that there is no diffeomorphism Ward identity. 

Such a theory is not conformally invariant, but there is a symmetry under scaling transformations provided that we also allow the coupling to transform i.e. there is invariance
under Weyl rescaling of the metric and other fields
\be
g_{ij} \rightarrow e^{2 \sigma} g_{ij} \qquad \cdots
\ee
provided that we also scale the coupling as 
\be
\Phi_s \rightarrow e^{-(d- \Delta_{\phi}) \sigma} \Phi_s.
\ee
It is this generalized conformal structure that is captured by the Ward identities, which imply an infinite set of relations for correlation functions. Associated with the (dimensionful) 
coupling $\phi_s$ is a dimensionless effective coupling 
\be
g_{\rm eff}^2 (x) = |x|^{\Delta_{\Phi} - d} \Phi^{-1}_s \label{eff-c}
\ee
which controls the strength of interactions. 

When ${\cal O}_{\Phi}$ is a relevant operator, the effective coupling becomes large in the IR and conversely the effective coupling becomes large in the UV when ${\cal O}_{\Phi}$ is an irrelevant operator. Typically the generalized conformal structure controls the dynamics over a range of energies but an alternative description is required when the effective coupling becomes too large. 

Now let us consider correlation functions of scalar operators in a theory with generalised conformal structure. A general scalar operator ${\cal O}$ has two point functions
\be
\langle {\cal O}(x) {\cal O}(0) \rangle = \frac{f(g_{\rm eff}^2 (x), N, \cdots )}{|x|^{2 \Delta}} \label{func}
\ee
where {\it any} function $f$ of the dimensionless quantities is consistent with generalized conformal structure. Generically, this function will be expressed as an analytic function of the dimensionless quantities but it can happen in certain limits that a single monomial in the analytic series dominates i.e. 
\be
f(g_{\rm eff}^2 (x), N, \cdots ) \sim |x|^{2 \alpha}
\ee
where $\alpha$ is a particular (not necessarily integral) power. 

The above structure may seem contrived and unlikely to be preserved at the quantum level. However, there are a number of concrete examples of theories exhibiting generalized complex structure. 

\subsection{Example 1: Maximal SYM theories}

The Euclidean action for maximal SYM in $d$ dimensions contains the following bosonic terms:
\be
I = - \int d^d x \left ( - \Phi_s {\rm Tr} (F_{ij} F^{ij}) + {\rm Tr} (D_i X D^i X) + \frac{1}{\Phi_s} {\rm Tr} [X,X]^2 + \cdots \right )
\ee
where $F_{ij}$ are the $SU(N)$ gauge fields and $X$ are the $(10-d)$ scalars. 

In this case the scalar operator ${\cal O}_{\Phi}$ associated with the generalized conformal structure is given by 
\be
{\cal O}_{\Phi} = {\rm Tr} (F_{ij} F^{ij}) + \frac{1}{\Phi_s^2} {\rm Tr} [X,X]^2
\ee
i.e. it is the gluon operator. 

In these theories the generalized conformal structure is believed to be respected at the quantum level, due to maximal supersymmetry. This can be shown
directly in perturbation theory and also by showing that the holographic duals at strong 't Hooft coupling respect the generalized conformal structure. As discussed above, 
the dimensionally running coupling implies that an alternative description is required either in the deep IR or UV. For the maximally SYM theories, the range of validity of  
the regime governed by generalized conformal structure is well understood in terms of decoupling limits of Dp-branes \cite{Itzhaki:1998dd} and the IR/UV completions within string theory are also 
known. For example, the 5d SYM associated with D4-branes is completed in the UV by the 6d M5-brane CFT. 

One can calculate the two point function of the gluon operator both at weak coupling, using a two loop calculation, and at strong coupling, using holography. 
For $g_{\rm eff}^2 N \rightarrow 0$ the two loop calculation gives \cite{Kanitscheider:2008kd}
\be
\langle {\cal O}_{\Phi} (x) {\cal O}_{\Phi} (0) \rangle \sim \frac{g_{\rm eff}^4 (x) N^2}{|x|^8}
\ee
while for $g_{\rm eff}^2 N \rightarrow \infty$ 
\be
\langle {\cal O}_{\Phi}(x) {\cal O}_{\Phi} (0) \rangle \sim \frac{g_{\rm eff}^{2 \frac{d-4}{6-d}}(x) N^2}{|x|^8}
\ee
from holography. Using the definition of the effective coupling \eqref{eff-c} and rescaling the operator by appropriate factors of $\Phi_s$ both results can be expressed as 
\be
\langle \tilde{\cal O}(x) \tilde{\cal O}(0) \rangle \sim |x|^{-2 \alpha}
\ee
for different values of $\alpha$. Thus if one only knows the correlator to leading order at weak or strong coupling one might assume that the scaling behaviour is indicative of an
underlying scale invariance - but away from these limits the function defined in \eqref{func} is a series, rather than a single term, reflecting the underlying generalized conformal 
structure. 

\subsection{Example 2: SYK}

From the relations written down earlier, it is clear that the SYK classical Lagrangian is associated with generalized conformal structure. We can write the (Euclidean) Lagrangian in a covariant manner as
\be
I = - \int d \tau \sqrt{g} \left ( -  \sum_i \psi^i \gamma^{\tau} \partial_{\tau} \psi^i  + \sum_{ijkl} \lambda_{ijkl} \psi^i \psi^j \psi^k \psi^l \right )
\ee
and from here derive as dilatation and diffeomorphism Ward identities the relations given earlier:
\be
{\cal H} + \sum_{ijkl} \lambda_{ijkl} \psi^i \psi^j \psi^k \psi^l  = 0 \qquad 
\partial_{\tau} {\cal H} + \sum_{ijkl} (\partial_{\tau} \lambda_{ijjkl} ) \psi^i \psi^j \psi^k \psi^l = 0.
\ee
These relations display generalized conformal structure, with many scalar operators rather than a single operator. For later use, it is convenient to denote the set of scalar operators and associated couplings as
\be
{\cal O}_A \equiv  \{ \psi^i \psi^j \psi^k \psi^l \} \qquad
\lambda_{A} \equiv \{ \lambda_{ijkl} \}. 
\ee
with the range of $A$ being of order $N^4$ in the large $N$ limit. 

Associated with each of these dimensionful couplings is a dimensionless coupling 
\be
(\tilde{\lambda}_{\rm eff})_{ijkl}(\tau)  = | \tau | \lambda_{ijkl}.
\ee
(It is convenient to define the dimensionless coupling as the inverse of the relation used for the gauge theories in the previous subsection.) 

Now consider the two point function of a fermion $\psi^i$, calculated perturbatively in $\lambda_{ijkl}$. Since a free fermion is dimensionless in one dimension, 
the two point function is logarithmic (as for free bosons in two dimensions). Let us therefore regulate by working in $d = 1 + \epsilon$ dimensions. Then the fermion two point function is given perturbatively by 
\be
\langle \psi^i(\tau) \psi^i (0) \rangle \sim \frac{1}{\tau^{\epsilon}} \left ( 1 + c \sum_{jkl} (\tilde{\lambda}_{\rm eff})^2_{ijkl}(\tau) + \cdots \right )
\ee
where $c$ is a numerical constant (of order $N^3)$ that can be calculated explicitly via two loop calculation of the diagrams shown in Figure~\ref{fig:one}. 

\FIGURE[t]{\includegraphics*{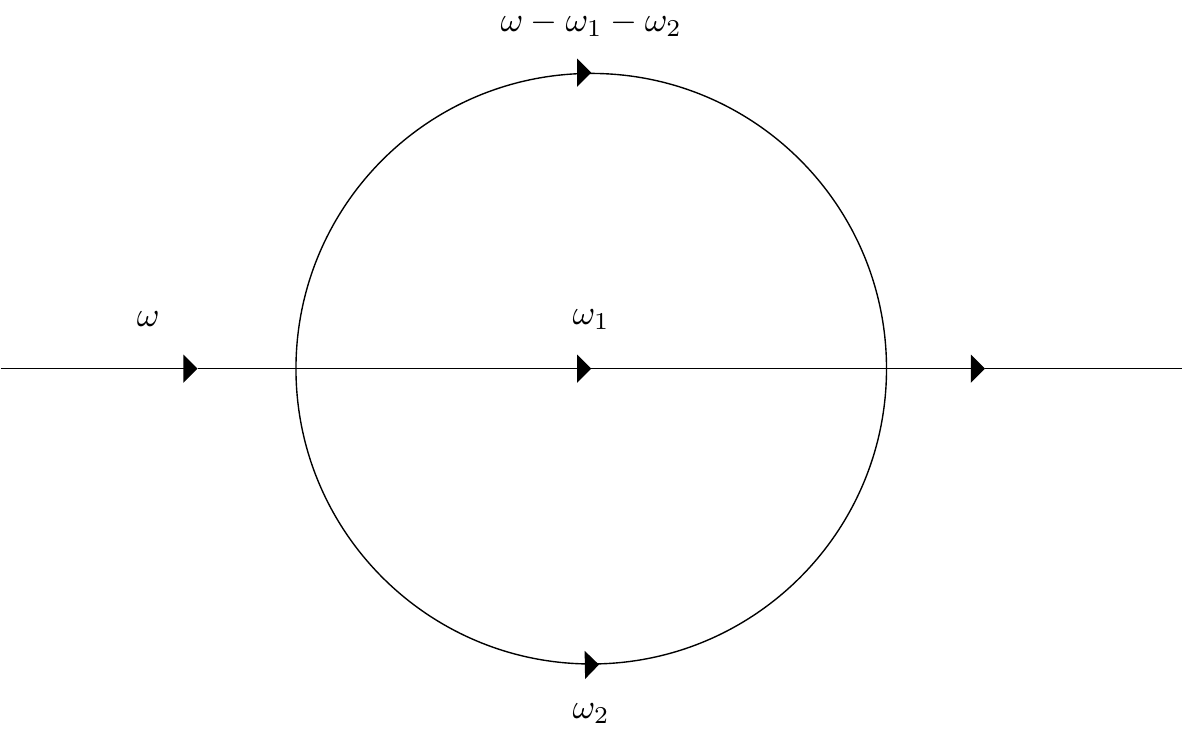}
\caption{Two loop contributions to the fermion two point function.}
\label{fig:one}}

The perturbative result is valid provided that 
\be
\sum_{jkl} {\lambda}^2_{ijkl} \tau^2 \ll 1.
\ee
In the SYK model, the Gaussian governing the random couplings has width ${\cal J}/N^{3/2}$ and hence 
\be
\sum_{jkl} {\lambda}^2_{ijkl} \sim {\cal J}^2
\ee
and thus the expression
\be
\langle G(\tau,0) \rangle \equiv \frac{1}{N} \sum_i \langle \psi^i(\tau) \psi^i (0) \rangle \sim \frac{1}{|\tau|^{\epsilon}} \left ( 1 + \tilde{c}  {\cal J}^2 |\tau|^2 + \cdots \right )
\ee
(where $\tilde{c}$ is of order one) is valid for small separations such that ${\cal J} \tau \ll 1$. This is the opposite limit to the results summarised in the previous section. 

\bigskip

Now let us review the behaviour of this model in the large $N$ and large ${\cal J}$ limit. Implementing the Gaussian averaging over the couplings
gives an effective action of the form 
\be
I_{\rm eff} = - \int \;  d\tau \psi_i \partial_\tau \psi_i - \frac{{\cal J}^2}{4 q} \int d\tau_1 d\tau_2 | \psi_i (\tau_1) \psi_i (\tau_2) |^q \label{eff2}
\ee
where implicitly we sum over the repeated indices. 

To solve the model in the large ${\cal J}$ limit, it is convenient to work in terms of the bilocal field $G(\tau_1,\tau_2)$ defined in \eqref{bilocal}, for which the action is
\be
I_{\rm bilocal} = - \int d\tau \left ( \partial_\tau G(\tau,\tau')_{\tau=\tau'}  + {\rm Tr} \left (\log G \right )  \right ) - \frac{ {\cal J}^2}{4} \int d\tau_1 d\tau_2 | G(\tau_1,\tau_2)|^q.
\ee
One can then rescale the bilocal field by appropriate factors of ${\cal J}$ such that only the first term in the action contains a dimensionful factor:
\be
I_{\rm bilocal} = - \int d\tau \left ( \frac{1}{{\cal J}^{\frac{2}{q}}} \partial_\tau \tilde{G}(\tau,\tau')_{\tau=\tau'}  + {\rm Tr} \left (\log \tilde{G} \right ) \right ) - \frac{1}{4} \int d\tau_1 d\tau_2 | \tilde{G}(\tau_1,\tau_2)|^q.
\ee
The latter two terms are manifestly scale invariant, with scale invariance broken explicitly by the scale dependence in the first term. 

This realisation of the action allows the correlation functions to be determined 
perturbatively for ${\cal J} |\tau | \gg 1$. In particular, as advertised earlier
\be
\langle \tilde{G}(\tau,0) \rangle = \frac{1}{|\tau|^{2/q}} + \frac{b}{{\cal J} |\tau|^{2/q + 1}} + \cdots
\ee
where the constant $b$ can be found in \cite{Maldacena:2016hyu}. This expression can clearly be written in a way to display generalized conformal structure as 
\be
\langle \tilde{G}(t,0) \rangle = \frac{f({\cal J} |\tau|)}{|\tau|^{2/q}},
\ee
where $f$ admits an expansion in ${\cal J} |\tau|$. Note that the dimension for $\tilde{G}(\tau,0)$ is as expected given the factors of ${\cal J}$ in the operator 
definition. 

Returning to \eqref{eff2}, this effective action clearly demonstrates generalised conformal structure with the following dilatation Ward identity:
\be
{\cal H} = \frac{ 1}{4 q} {\cal O}
\ee
where
\be
\int {\cal O}(\tau) d\tau = \int d\tau_1 d\tau_2 | \tilde{G}(\tau_1,\tau_2)) |^{q}
\ee
Thus, the effect of the random averaging is that a {single} effective operator controls the dimensionally driven flow.  We will discuss this interpretation further later.

\section{Holographic realization of generalized conformal structure} \label{four}


Consider the following Euclidean dilaton gravity action in $(d+1)$ dimensions:
\be
I = - {\cal N} \int d^{d+1} x \sqrt{g} e^{\gamma \phi} \left ( R + \beta (\partial \phi)^2 + C \right ). \label{ingmar0}
\ee
Such actions arise in the context of non-conformal brane holography. In AdS holography, Einstein gravity with negative cosmological constant describes the
energy momentum tensor of the dual theory; it arises as a consistent truncation of the top down supergravity theory compactified on a sphere. In non-conformal 
brane holography, Einstein gravity with a dilaton captures the dual energy momentum tensor as well as the scalar operator driving the flow; dilaton gravity again 
arises as a consistent truncation of the top down supergravity theory compactified on a sphere \cite{Boonstra:1998mp}. 

We will discuss the normalization of the action ${\cal N}$ in section \ref{seven}. The action is expressed in the dual frame, which was first introduced in \cite{Duff:1994fg};
in this frame the equations of motion admit solutions in which the metric is anti-de Sitter. The dual frame is the natural frame in which to set up the holographic dictionary \cite{Kanitscheider:2008kd}. 

The equations of motion following from \eqref{ingmar0} admit $AdS_{d+1}$ solutions with running dilaton:
\bea
ds^2 & = & \frac{1}{\rho^2} \left ( d \rho^2 + dx \cdot dx_d \right )\\
e^{\phi} &=& \rho^{2 \alpha}  \nn
\eea
where 
\bea
\alpha &=& - \frac{\gamma}{2 (\gamma^2 - \beta)} \\
C &=& \frac{ (d (\gamma^2 - \beta) + \gamma^2)( d (\gamma^2 - \beta) + \beta)}{(\gamma^2 - \beta)^2}. \nn
\eea
To understand how the dual field theory data is encoded in the bulk asymptotics, we consider generally asymptotically locally
AdS solutions of the form
\bea
ds^2 &=& \frac{1}{\rho^2} \left ( d \rho^2 + g_{ij} (x,\rho) dx^i dx^j \right ) \\
\phi &=& 2 \alpha \log (\rho) + \kappa (x, \rho) \nn
\eea
where 
\bea
g_{ij} (x,\rho) &=& g_{(0) ij} (x) + g_{(2) ij} (x) \rho^2 + \cdots \\
\kappa(x, \rho) &=& \phi_s(x) +  \kappa_{(2)} \rho^2 + \cdots \nn
\eea
Here $g_{(0)ij}$ is the background metric for the dual theory and $\phi_s(x)$ is the coupling in the field theory. One can show using holographic 
renormalization \cite{Kanitscheider:2008kd} that the operators
$T_{ij}$ and ${\cal O}_{\phi}$ dual to the metric and scalar 
take expectation values that  can be expressed as terms in the asymptotic expansion:
\bea
\langle {\cal T}_{ij}  \rangle &=& (d - 2 \alpha \gamma) {\cal N} e^{\phi_s} g_{(2 \sigma) \tau \tau} + \cdots \\
\langle {\cal O}_{\phi} \rangle &=& 2 (d- 2\alpha \gamma) {\cal N} e^{\phi_s } \frac{\kappa_{(2 \sigma)}}{\alpha} + \cdots \nn
\eea
where the ellipses denote additional contributions that arise when when $(d - 2 \alpha \gamma)$ is an even integer. 

These operator expectation values satisfy the following dilatation and diffeomorphism Ward identities:
\bea
\langle T^{i}_{i} \rangle + 2 \alpha \langle {\cal O}_{\phi} \rangle &=& {\cal A}, \label{wi-2} \\
\nabla^i \langle T_{ij} \rangle - \partial_j \phi_s \langle {\cal O}_{\phi} \rangle &=& 0, \nn
\eea
where the anomalies ${\cal A}$ that arise when $(d - 2 \alpha \gamma)$ is an even integer are discussed in detail in \cite{Kanitscheider:2008kd}. 

\bigskip

Let us now consider the specific case of non-conformal Dp-branes for which one can show that 
\be
\alpha = - \frac{ (p-7) (p-3)}{4 (p-5)} \qquad \gamma = \frac{2 (p-3)}{(7-p)}.
\ee
The bulk scalar field is related to the operator defined in the previous section as follows. Let $\Phi$ be the bulk field coupling to this operator. This field is related to the field $\phi$ as 
\be
\Phi(x,\rho) = \exp \left ( \chi \phi(x,\rho) \right ) \label{rescale}
\ee
where 
\be
\chi = \frac{2 (p-5)}{(p-7)}.
\ee
(Note that the above expressions do not apply for $p=5,7$; in any case a decoupling regime for which the dynamics is governed by generalized conformal structure 
exists only for $p < 5$.) The asymptotic expansion of $\Phi$ is
\be
\Phi(x,\rho) = \rho^{- (p-3)} \left ( \Phi_s(x) + \cdots \right )
\ee
where $\Phi_s(x)$ is given by 
\be
\Phi_s (x) = \exp \left (  \frac{2 (p-5)}{(p-7)} \phi_s(x) \right )
\ee
i.e. $\Phi_s(x) = 1$ when $\phi_s (x) = 0$. Using the relation $\langle {\cal O}_{\phi} \rangle = \chi \Phi_s \langle {\cal O}_{\Phi} \rangle $ we can then write the Ward identities as
\bea
\langle T^{i}_{i} \rangle + (p-3) \Phi_s \langle {\cal O}_{\Phi} \rangle &=& {\cal A}, \\
\nabla^i \langle T_{ij} \rangle + \partial_j \Phi_s \langle {\cal O}_{\Phi} \rangle &=& 0, \nn
\eea
i.e. the same identities as discussed in the previous section. 

It is known \cite{Kanitscheider:2009as} that the dilaton gravity action can always be interpreted as the reduction of an AdS gravity theory in $(2 \sigma + 1)$ dimensions
where 
\be
2 \sigma = d - 2 \alpha \gamma
\ee
(where $2 \sigma$ is not necessarily an integer)
and the reduction ansatz is over a $(2 \sigma - d)$-dimensional torus 
\be
ds^2 = ds_{d+1}^2 + \exp (2 \gamma \phi) dz \cdot dz_{(2 \sigma - d)}. \label{ingmar2}
\ee
As we will discuss later, the parameter $\sigma$ controls the thermodynamics and structure of the correlation functions for the theory, e.g. the powers of $|x|$ appearing in the operator two point functions. Thus, while we can always rescale the dual operator (by dimensional factors) as in \eqref{rescale}, the parameters appearing in the holographic realization have important physical consequences. 

\subsection{AdS$_3$ reduction}

A particular case of interest in the context of nearly AdS$_2$ (NAdS$_{2}$) is  $\beta =0$, so that
\be
2 \alpha \gamma = -1 \qquad C = d (d+1). 
\ee
In the NAdS$_2$ case, the reduction is over a circle from $AdS_3$. Since the scalar kinetic term vanishes, the scalar can always be rescaled; it is convenient to choose
this rescaling such that $\gamma = 1$. 

There is an immediate generalization \cite{Gouteraux:2011qh} to a reduction including a Kaluza-Klein gauge field $A_{\mu}$:
\be
ds^2 = ds_{d+1}^2 + \exp(2 \gamma \phi) (dy + A_{\mu} dx^{\mu})^2
\ee
with the resulting reduced action being 
\be
I = - {\cal N} \int d^{d+1} x \sqrt{g} e^{\gamma \phi} \left ( R + d (d+1) - \frac{1}{4} e^{2 \gamma \phi} F_{\mu \nu} F^{\mu \nu} \right ).
\ee
Again the vanishing of the scalar kinetic term implies that we can set $\gamma = 1$. For $d=1$, note that we cannot reach an Einstein frame 
by rescaling the metric by conformal factors as these factors cancel in the first term of the action. The AdS$_3$ parent theory to SYK was discussed 
\cite{Cvetic:2016eiv} and the holographic dictionary was derived in detail in this work. 

\bigskip

The holographic dictionary of the dilaton gravity theory follows immediately from the standard holographic dictionary for $AdS_3$. In the latter case,
the bulk Einstein theory captures the dual stress energy tensor ${\cal T}_{ij}$. As usual, the expectation value of the latter follows from expanding the asymptotically locally $AdS_3$ metric near the conformal boundary as
\be
ds^2 = \frac{d \rho^2}{\rho^2} + \frac{1}{\rho^2} \left ( g_{(0) ij} + \rho^2 (g_{(2) ij} + \ln \rho h_{(2) ij} + \cdots \right ) dx^i dx^j
\ee
Here $g_{(0) ij}$ is the background metric for the dual theory and 
\be
\langle {\cal T}_{ij} \rangle = \frac{c}{24 \pi} \left ( g_{(2) ij} + \frac{1}{2} {\cal R} g_{(0) ij} \right )
\ee
where ${\cal R}$ is the scalar curvature of $g_{(0)ij}$ and the central charge $c$ is related to the normalisation of the bulk action as $c = 24 \pi {\cal N}$.

The Ward identities for the stress energy tensor are 
\be 
\langle {\cal T}_{i}^i \rangle = \frac{c}{6} {\cal R} \qquad 
\nabla^i \langle {\cal T}_{ij} \rangle = 0,
\ee
where the covariant derivative is constructed from the metric $g_{(0) ij}$. 

\bigskip

On dimensional reduction, we decompose the stress energy tensor in Fourier modes as 
\be
{\cal T}_{ij} (\tau,y) = \sum_{n \ge 0} {\cal T}^{(n)}_{ij} (\tau) e^{ i n y}
\ee
and we denote ${\cal T}^{(0)}_{\tau \tau} \equiv {\cal H}$, ${\cal T}^{(0)}_{yy} \equiv {\cal O}$ and ${\cal T}^{(0)}_{ \tau y} \equiv {\cal J}$. Then (in Euclidean Signature) the $n=0$
component of the Ward identities gives
\be
\langle {\cal H} \rangle + \langle {\cal O} \rangle = {\cal A};
\qquad \nabla_{\tau}  \langle {\cal H} \rangle = \nabla_{\tau}\langle {\cal O} \rangle = 0.  \label{1dWard}
\ee
If we express the background metric for the field theory as 
\be
ds^2 = \exp(2 a(\tau)) dt^2 + \exp ( 2 b(\tau)) \left ( dy + A(\tau) dt \right )^2
\ee
then the scalar curvature is 
\be
{\cal R} = 2 \exp(- 2 a (\tau)) \left ( \partial_{\tau} b ( \partial_{\tau} a - \partial_{\tau} b) - \partial_{\tau}^2 b \right ),
\ee
and the reduced anomaly ${\cal A} = c {\cal R}/6$.
(Since the off-diagonal term is locally pure gauge, one would not expect $A(t)$ to appear in the scalar curvature.) The inverse metric components
\bea
g_{(0)}^{\tau \tau} &=& \exp(-2 a(\tau)) \qquad
g_{(0)}^{\tau y} = -  \exp(-2 a(\tau)) A(\tau) \\\ 
g_{(0)}^{yy} &=& \exp(-2 b(\tau)) + \exp(-2 a(\tau) ) A(\tau)^2 \nn
\eea
gives the sources for the operators $({\cal H}, {\cal J}, {\cal O})$ respectively. 

\bigskip

Let us next consider the behaviour of a test scalar field $\varphi$ in $AdS_3$ dual to an operator ${\cal O}_{\varphi}$. As usual if the bulk field has mass $m^2$ then the corresponding conformal dimension $\Delta$ is such that $m^2 = \Delta ( \Delta - 2)$ and the (Euclidean) two point function of the operator in the conformal vacuum is 
\be
\langle {\cal O}_{\varphi} (\tau,y), {\cal O}_{\varphi} (0,0) \rangle = \frac{c_{\varphi}}{ (\tau^2 + y^2)^{\Delta}},
\ee
where we assume that the correlator is at separated points and $c_{\varphi}$ is the normalisation. Now let us Fourier transform along the $y$ direction 
(assuming that $y$ is infinite and $\tau$ is always finite). Thus, we write 
\be
{\cal O}_{\varphi} (\tau,k) = \int dy e^{iky} {\cal O}_{\varphi} (\tau,y)
\ee
and the correlator in this mixed representation is (for $k \ge 0$)
\be
\langle {\cal O}_{\varphi} (\tau,k) {\cal O}_{\varphi} (0,-k) \rangle \sim | \tau |^{1 - 2\Delta} e^{ - k | \tau |},
\ee
i.e. there is an exponential fall-off for non-zero momenta. 

This is precisely the same behaviour captured by the dynamics of the bulk scalar field. The field equation in pure $AdS_3$ is 
\be
\rho^3 \partial_{\rho} \left ( \frac{\partial_{\rho} \varphi}{\rho} \right ) + \rho^2 (\partial_{\tau}^2 + \partial_y^2) \varphi = m^2 \varphi.
\ee
Fourier transforming along the $y$ direction gives
\be
\rho^3 \partial_{\rho} \left ( \frac{\partial_{\rho} \varphi}{\rho} \right ) +  \rho^2 \partial_{\tau}^2 \varphi = (m^2 + k ^2 \rho^2) \varphi.
\ee
The momentum along the $y$ direction translates into a position dependent contribution to the mass: this contribution vanishes near the conformal boundary $\rho \rightarrow 0$ (UV of field theory) but
increases in the bulk interior and diverges as $\rho \rightarrow \infty$ (IR of field theory). This concurs with the UV limit of the correlator being power law while at larger time separations the fall-off is exponential. 

Now let us take the length of the $y$ direction to be $2 \pi$, so that the momentum is quantised in integral units. The corresponding modes of the scalar operator are expressed as 
\be
{\cal O}_{\varphi} (t,y) = \sum_{n \ge 0}^{\infty} {\cal O}^{(n)}_{\varphi} (\tau) e^{ in y}
\ee
and only the $n=0$ operator exhibits scaling behaviour in its two-point function:
\be
\langle {\cal O}^{(0)}_{\varphi} (\tau) {\cal O}^{(0)}_{\varphi} (0) \rangle \sim | \tau |^{1 - 2\Delta} 
\ee
From the bulk respective, the reduction over the circle (setting the KK gauge field to zero) 
gives the following actions for the scalar field modes $\varphi^{(n})(\rho,\tau)$ dual to ${\cal O}^{(n)}$:
\be
I =  \frac{1}{2} \int d^2 x \sqrt{g} e^{\gamma \phi} \left ( (\partial \varphi^{(n)})^2 + (m^2 + n^2 e^{-2 \gamma \phi}) (\varphi^{(n)})^2 \right ). 
\ee
Note that even the zero mode is not minimally coupled, due to the coupling to the running scalar $\phi$. 

As in AdS, such a field in NAdS$_2$ is dual to an operator that exhibits scaling behaviour but the relation between mass and scaling dimension is inherited from the upstairs picture rather than following the usual AdS$_2$ relation i.e. $m^2 = \Delta (\Delta - 2)$ in NAdS$_2$ as opposed to $m^2 = \Delta (\Delta  - 1)$ in AdS$_2$.

\bigskip

Now let us move from the test scalar example to the case of the energy momentum tensor. Two point functions of the energy momentum tensor in a two-dimensional CFT are more conventionally expressed in chiral and anti-chiral components. It is however straightforward to write these (Euclidean) correlators in Cartesian coordinates as 
\bea
\langle T_{\tau \tau} (\tau,y) T_{\tau \tau} (0) \rangle &=& \langle T_{yy}(\tau,y) T_{yy} (0) \rangle = \frac{c}{(2 \pi)^2} \frac{ (\tau^4 - 6 \tau^2 y^2 + y^4)}{(\tau^2 + y^2)^4} \\
\langle T_{\tau y} (t,y) T_{\tau \tau} (0) \rangle &=& \langle T_{yy} (\tau,y) T_{yy} (0) \rangle = \frac{4c}{(2 \pi)^2} \frac{ \tau y (\tau^2 - y^2)}{(\tau^2 + y^2)^4} \nn \\
\langle T_{\tau \tau} (\tau,y) T_{yy} (0) \rangle &=& \langle T_{\tau y}(\tau,y) T_{\tau y} (0) \rangle  = - \frac{c}{(2 \pi)^2} \frac{ (\tau^4 - 6 \tau^2 y^2 + y^4)}{(\tau^2 + y^2)^4}  \nn 
\eea
where we have suppressed scheme dependent local terms. We can now Fourier transform the correlation functions along the $y$ direction to determine the correlation functions of the Fourier modes of the operators. Using the standard integrals
\bea
\int^{\infty}_{- \infty} \frac{dy}{(\tau^2 + y^2)^4} &=& \frac{5 \pi}{16} \\ 
\int^{\infty}_{-\infty} \frac{ y^2 dy}{(\tau^2 + y^2)^4} &=& \int^{\infty}_{-\infty} \frac{ y^4 dy}{(\tau^2 + y^2)^4} = \frac{\pi}{16} \nn
\eea
we can then show that all of the two point functions between $({\cal H}, {\cal O}, {\cal J})$ actually vanish, up to local terms. As in the scalar operator example above, higher Fourier modes of the upstairs operators have exponentially decaying correlators:
\be
\langle {\cal T}^{(n)} (t) {\cal T}^{(- n)} (0) \rangle \sim \frac{e^{- n |\tau|} }{|\tau|^3}, 
\ee
and so on. 

This result follows directly from the diffeomorphism Ward identities. In 2d the correlators of the energy momentum tensor are calculating by integrating the Ward identities. Reducing to one dimension, we noted in \eqref{1dWard} that the diffeomorphism Ward identities imply that the operators are independent of time and this in turn implies that the two point functions are trivial. 

\bigskip

One can understand the results above using the more standard expressions of 2d CFT two-point functions in chiral coordinates. An operator of ${\cal O}_{p,q}$ with scaling weights $(p,q)$ has a two-point function
\be
\langle {\cal O}_{p,q} (z, \bar{z}) {\cal O}_{p,q} (0,0) \rangle = \frac{c_{o}}{ z^{2p} \bar{z}^{2q}}
\ee
where $z = \tau + i y$. Fourier transforming with respect to the $y$ direction thus requires the integrals
\be
\int^{\infty}_{- \infty} \frac{dy}{(\tau + i y)^{2p} (\tau - i y)^{2q}} e^{- i n y}
\ee
with the relevant modes for the reduced theory being the $n=0$ terms. In the latter case the integral can be calculated to give
\be
\langle {\cal O}^{(0)}_{p,q} (t) {\cal O}^{(0)}_{p,q} (0) \rangle = \frac{\pi (2 q + 2p -2)! }{2 ^{2 (p+q-1)} (2p-1)!(2q-1)! } \frac{c_o}{\tau^{2 p + 2q -1}} \label{red2pf}
\ee
where for simplicity we restrict to $(p,q)$ integral and as above we decompose the operator as 
\be
{\cal O}_{p,q} (z, \bar{z}) = \sum_{n} {\cal O}_{p,q}^{(n)} (\tau) e^{ i n y}.
\ee
The expression \eqref{red2pf} vanishes if either $p =0$ with $q \neq 0$ or $q =0$ with $p \neq 0$. For marginal $(1,1)$ operators 
\be
\langle {\cal O}^{(0)}_{1,1} (t) {\cal O}^{(0)}_{1,1} (0) \rangle =  \frac{\pi c_o}{4 \tau^{3}}
\ee
However for the stress energy tensor expressed in the usual chiral coordinates the correlators
\be
\langle {\cal T}_{zz} (z) {\cal T}_{zz} (0) \rangle = \frac{c_L}{2 z^4} \qquad
\langle {\cal T}_{\bar{z} \bar{z}} (\bar{z}) {\cal T}_{\bar{z} \bar{z}}(0) \rangle = \frac{c_R}{2 \bar{z}^4} 
\ee
reduced to the $n=0$ modes give zero as they correspond to $(2,0)$ and $(0,2)$ operators. 
Since the operators $({\cal H}, {\cal J}, {\cal O})$ defined above can be expressed as linear combinations of these operators, this explains
why their correlators were also zero. 

For general dimensions $d > 1$, the parameter $\sigma$ controls the scaling behavior of the two point functions of the energy momentum tensor and scalar operator.
When the dual theory is one dimensional these correlators are forced by the Ward identities to vanish. 
The parameter $\sigma$ however still has physical relevance as it controls the thermodynamic properties.

\subsection{Finite temperature behaviour}

We now briefly review the reduction of black hole solutions of the 3d theory, see \cite{Cvetic:2016eiv} for more details. 
It is convenient to write the rotating BTZ black hole in Fefferman-Graham coordinates as 
\bea
ds^2 &=& \frac{ d \rho^2}{\rho^2} - \left ( \frac{1}{\rho^2} - \frac{1}{2} ( r_{-}^2 + r_{+}^2 ) + \frac{1}{4} (r_{+}^2 - r_{-}^2)^2 \rho^2 \right ) dt^2 \\
&& \qquad 
+ \left ( \frac{1}{\rho^2} + \frac{1}{2} (r_{+}^2 + r_{-}^2) + \frac{1}{4} (r_{+}^2 - r_{-}^2 )^2 \rho^2 \right ) d y^2 + 2 r_{+} r_{-} dt d y \nn
\eea
Here we work in Lorentzian signature and correspondingly use the coordinate $t$ for the Lorentzian time. 

Upon dimensional reduction, the two-dimensional fields are:
\bea
ds^2 &=& - \frac{(r^2 - r_{+}^2)(r^2 - r_{-}^2)}{r^2} dt^2 + \frac{r^2 dr^2}{(r^2 - r_{+}^2)(r^2 - r_{-}^2)} \\
A_t &=& \frac{r_{+} r_{-}}{r^2} \qquad 
e^{\phi} = r \nn
\eea 
Evaluating the one-point function of the dual 2d energy momentum tensor gives
\bea
\langle {\cal T}_{tt} \rangle &=& \langle {\cal T}_{yy} \rangle = {\cal N} (r_{+}^2 + r_{-}^2 ) \\
\langle {\cal T}_{t y} \rangle &=& {\cal N} (r_{+} r_{-}). \nn
\eea
Dimensionally reducing, only the zero mode components of the operators (wrt the $y$ circle) have non-vanishing expectation values, i.e. 
\bea
\langle {\cal H} \rangle &=& \langle {\cal O} \rangle = {\cal N}(r_{+}^2 + r_{-}^2) \\
\langle {\cal J} \rangle &=& {\cal N} ( r_{+} r_{-}). \nn
\eea
These satisfy the required Lorentzian Ward identities.

Redefining the radial coordinate as 
\be
r^2 = \frac{1}{\rho^2} + \frac{1}{2} (r_{+}^2 + r_-^2) + \frac{1}{4} (r_{+}^2 - r_{-}^2)^2 \rho^2
\ee
the metric can be written in the more familiar form
\be
ds^2 = - (r^2 - r_{+}^2 - r_{-}^2) dt^2 + r^2 d y^2 + 2 r_{+} r_{-} dt dy + \frac{r^2 dr^2}{(r^2 - r_{+}^2)(r^2 - r_{-}^2)}
\ee 
which makes manifest that the horizons are located at $r = r_{\pm}$. The temperature is given by
\be
T_{H} = \frac{r_{+}^2 - r_{-}^2}{2 \pi r_{+}}.
\ee
and the entropy is 
\be
S = 4 \pi {\cal N} r_{+}.
\ee
In the extremal limit $r_{+} \rightarrow r_{-}$ and $T_{H} \rightarrow 0$ with $S$ finite. Thus the operators acquire expectation values in the thermal state and 
the thermodynamic properties agree with those of SYK. 

\section{Holographic realization of generalized conformal structure with multiple scalar fields} \label{five}

In this section we will consider the holographic realization of generalized conformal structure, for flows driven by multiple scalar operators. The motivation is SYK, but the construction is more generally applicable. For brevity we analyze the case of two bulk dimensions but similar constructions could be made in all dimensions. 

Consider a 2d bulk gravity action with scalar fields $\phi_A$:
\be
I = - {\cal N} \int d^{2} x \sqrt{g} \; e^{\sum_{A} \gamma_A \phi_A} \left ( R + \sum_{B,C} \beta_{BC} (\partial \phi_B)(\partial \phi_C) + C \right ). \label{mult-act}
\ee
The scalar field equations of motion are
\be
\gamma_A \left (R + \sum_{B,C} \beta_{BC} (\partial \phi_B)(\partial \phi_C) + C \right ) =  \sum_C \beta_{AC} \left ( \nabla^2 \phi_C +  \partial_{\mu} \phi_C \sum_B \gamma_B (\partial^{\mu} \phi_B) \right ).
\ee
The Einstein equations are
\bea
&& - R_{\mu \nu} - \sum_{A,B} \beta_{AB} \partial_{\mu} \phi_A \partial_{\nu} \phi_B + \frac{1}{2} g_{\mu \nu} \left (R + \sum_{A,B} \beta_{AB} (\partial \phi_A)(\partial \phi_B) + C \right ) + \sum_A \gamma_A \nabla_{\mu} \partial_{\nu} \phi_A  \nn \\
&& \qquad + \sum_{A,B} \gamma_A \gamma_B \partial_{\mu} \phi_A \partial_{\nu} \phi_B - g_{\mu \nu} 
\left ( \sum_{A,B} \gamma_A \gamma_B (\partial^{\rho}\phi_A) (\partial_{\rho} \phi_B) + \sum_A \gamma_A \nabla^2 \phi_A \right ) = 0.
\eea
For $\beta_{AB} = 0$ the equations admit a solution
\be
ds^2 = \frac{d \rho^2}{\rho^2} + \frac{d \tau^2}{\rho^2} \qquad
e^{\phi_A} = {\rho^{\alpha_A}} \label{sol1}
\ee 
 provided that $\sum_A \alpha_A = - 1$. (We can trivially rescale the scalar fields so that $\gamma_A = 1$.) In this case we could simply define a single scalar $\phi \equiv 
 \sum_A \alpha_A \phi_A$, if we restrict to only the gravity/dilaton sector, but this would not be possible if the full theory has additional fields (such as gauge fields) that couple to the individual scalars differently. 
 
Solutions of the form \eqref{sol1} can be found for general values of $\alpha_A$ provided that $\beta_{AB}$ and $C$ are chosen appropriately. We can always rescale the scalars such that $\gamma_A  = 1$ and then the scalar field equations are satisfied provided
\be
(C - 2  + \sum_{B,C} \beta_{BC} \alpha_B \alpha_C) = 2 \sum_{C} \beta_{AC} \alpha_{C} (\sum_{B} \alpha_B  - 1).
\ee
The Einstein equations then imposes
\be
C + \sum_{B,C} \beta_{BC} \alpha_B \alpha_C = 2 (\sum_A \alpha_A)^2
\ee
and
\be
C = \sum_{B,C} \beta_{BC} \alpha_B \alpha_C - 2 \sum_{\alpha} \alpha_A
\ee 
Note that these equations are not linearly independent, as they are related by the Bianchi identity.
Combining the latter two equations gives
\be
C = \sum_{A}  \alpha_A ( \sum_{B} \alpha_B-1).
\ee
All equations are then satisfied if 
\be
\sum_{B} \beta_{AB} \alpha_B = \sum_{C} \alpha_C  + 1
\ee
which can be solved by the diagonal matrix:
\be
\beta_{AB} = \frac{1}{\alpha_A} ( \sum_{C} \alpha_C  + 1) \delta_{AB}.
\ee
We could have chosen $\beta_{AB}$ to be a diagonal matrix from the start, but again if the full theory has additional fields such as gauge fields with fixed couplings to the scalars  it is then not always possible to diagonalize the scalar kinetic terms; see discussions in \cite{Gouteraux:2011qh}.
 
\bigskip 
 
To understand how the dual field theory data is encoded in the bulk asymptotics, we now consider asymptotically locally
AdS$_2$ solutions in radial gauge i.e. 
\bea
ds^2 &=& \frac{1}{\rho^2} \left ( d \rho^2 + g_{\tau \tau} (\tau,\rho) d \tau^2 \right ) \\
\phi_A  &=& \alpha_A \log (\rho) + \kappa_A (\tau, \rho) \nn
\eea
where 
\bea
g_{\tau \tau} (\tau,\rho) &=& g_{(0) \tau \tau} (\tau) +  \cdots + g_{(2 \sigma)} \rho^{2 \sigma} + \cdots \\
\kappa_A(\tau, \rho) &=& \kappa_A (\tau) + \cdots + \kappa_{( 2\sigma) A} \rho^{2 \sigma} + \cdots \nn
\eea
and the powers arising in the expansion are determined by solving the field equations near the conformal boundary. In particular, in analogy to the single scalar case discussed
previously, the undetermined (normalizable) terms in the expansion arise at powers 
\be
2 \sigma = (1  - \sum_{A} \alpha_A ) \label{sig-def}
\ee
which implies that the asymptotic expansion makes sense only if $\sum_A \alpha_A < 1$ so that $\sigma > 0$. 

Here $g_{(0) \tau \tau}$ is the source for the Hamiltonian ${\cal H}$ and $\kappa_A(x)$ are the couplings for the operators ${\cal O}_{\phi_A}$ in the dual theory. Following the same steps as in \cite{Kanitscheider:2008kd} one can show that the operator  expectation values can be expressed as terms in the asymptotic expansion:
\bea
\langle {\cal H} \rangle &=& 2 \sigma {\cal N} e^{\sum_A \kappa_A} g_{(2 \sigma) \tau \tau} + \cdots \\
\langle {\cal O}_{\phi_A} \rangle &=& 4 \sigma {\cal N} e^{\sum_A \kappa_A} \frac{\kappa_{(2 \sigma)A}}{\alpha_A} + \cdots \nn
\eea
where the ellipses denote additional contributions when $\sigma$ is an integer. 

The operators manifestly satisfy the following dilatation and diffeomorphism Ward identities:
\bea
\langle {\cal H} \rangle + \sum_A \alpha_A \langle {\cal O}_{\phi_A} \rangle &=& {\cal A}, \\
\partial^{\tau} \langle {\cal H} \rangle - \sum_A \partial_{\tau} \kappa_A \langle {\cal O}_{\phi_A} \rangle &=& 0. \nn
\eea
Here again anomalous terms can arise when $\sigma$ is an integer. 

\bigskip

As for the non-conformal Dp-brane discussion, we can rescale the dual operator: let
\be
\Phi_A (\tau, \rho)= \exp (\chi_A \phi_A (\tau,\rho))
\ee
so that the asymptotic expansion of $\Phi_A$ is
\be
\Phi_A (\tau, \rho) = \rho^{\alpha_A \chi_A} \left ( \lambda_{A} (\tau) + \cdots \right ) \qquad 
\lambda_A (\tau) = e^{\chi_A \kappa_A (\tau)}. \label{asym-2}
\ee 
In terms of these rescaled operators the Ward identities become
\bea
\langle {\cal H} \rangle + \sum_A \alpha_A \chi_A \lambda_A \langle {\cal O}_{\Phi_A} \rangle &=& {\cal A}, \\
\partial_{\tau} \langle {\cal H} \rangle - \sum_A \partial_{\tau} \lambda_A \langle {\cal O}_{\Phi_A} \rangle &=& 0. \nn
\eea
This identity can be interpreted as generalized conformal structure where
\be
\alpha_A \chi_A = ( 1 - \Delta_A)
\ee
where $\Delta_A$ is the dimension of the operator ${\cal O}_{\Phi_A}$.  

\bigskip

In \cite{Kanitscheider:2009as} it was shown that holographic non-conformal theories can be described in terms of AdS gravity theories in $(2 \sigma + 1)$ 
dimensions. The arguments of \cite{Kanitscheider:2009as} can be extended to the case discussed above: the action \eqref{mult-act} can be obtained as the 
(formal) reduction of AdS gravity in $(2\sigma + 1)$ dimensions 
\be
I \propto -  \int d^{2 \sigma + 1} x \sqrt{G} \left ( R  +  2 \sigma (2 \sigma + 1) \right ) \label{ads-up}
\ee
using the reduction ansatz
\be
 ds^2 = g_{\mu \nu} dx^{\mu} dx^{\nu} + \sum_A e^{2 \Phi_A} dy_A^2.
 \ee
 Here the dimension of each direction $y_A$ is $ -\alpha_A$, i.e. non-integral in general, 
 so the total dimension of the compactification torus is $- \sum_A \alpha_A$.

 \subsection{Black hole solutions}
 
The equations of motion following from the action \eqref{mult-act} admit the following (neutral) black hole solutions
\bea
ds^2 &=& \frac{dr^2}{r^2 \left ( 1 - \frac{m}{r^{2 \sigma}} \right )} - r^2 \left ( 1 - \frac{m}{r^{2 \sigma}} \right ) d t^2 \\ 
e^{\phi_A} &=& \frac{1}{r^{\alpha_A}}. \nn
\eea 
These solutions follow immediately from the standard AdS black hole solutions in $(2 \sigma + 1)$ dimensions, i.e. black hole
solutions to the equations of motion following from \eqref{ads-up}. 

The temperature of the black hole is
\be
T_{H} = \frac{\sigma m^{\frac{1}{2 \sigma}}}{2 \pi}. \label{temp2}
\ee
The operator expectation values are
\be
\langle {\cal H} \rangle = {\cal N} (2 \sigma - 1) m
\qquad  \langle {\cal O}_{\phi_A} \rangle = -  {\cal N} m
\ee
 which indeed satisfy the (Lorentzian) Ward identity $\langle {\cal H} \rangle - \sum_A \alpha_A \langle {\cal O}_{\phi_A} \rangle \sim 0$  using \eqref{sig-def}. 
 The associated black hole entropy follows from the first law of thermodynamics as
 \be
 S = 4 \pi {\cal N} m^{1 - \frac{1}{2\sigma}}.
 \ee
Thus
\be
I_E = - 2\pi {\cal N} m^{1 - \frac{1}{2\sigma}}
\ee
This expression can be rewritten in terms of the temperature as 
\be
I_E \propto T_H^{2\sigma -1}
\ee
i.e. as we stated earlier the parameter $\sigma$ controls the thermodynamic properties of the theory: the thermodynamics is that of a conformal field theory 
in $2 \sigma$ dimensions.  

Thus, if one requires that the partition function scale linearly with the temperature, the parameter $\sigma$ must be equal to one. However, this case corresponds to
$\beta_{AB} = 0$, i.e. vanishing kinetic terms for the scalars, for which the effect of all of the scalars on the background can be captured by a single scalar
 $\phi = \sum_A \phi_A$. We will return to this point below. 

\subsection{Example: AdS$_5$ reduced on $T^3$}

To illustrate the general discussions, let us consider the case of asymptotically AdS$_5$ solutions reduced on $T^3$. The five-dimensional action is 
\be
I = - {\cal N}_5 \int d^4 x \sqrt{G} \left ( R(G) + 20 \right )
\ee
where ${\cal N}_5 \propto N^{2}$ for $AdS_5$ solutions dual to ${\cal N} = 4$ SYM, and variants thereof.  
The reduction ansatz is
\be
ds^2 = ds_2^2 + e^{2 \phi_1} dy_1^2 + e^{2 \phi_2} dy_2^2 + e^{2 \phi_3} dy_3^2
\ee
(where for simplicity we set the KK gauge fields to zero). The equations of motion for the reduced 2d theory then follow from the action
\be
I = - {\cal N} \int d^2 x \sqrt{g} e^{ \phi_1 + \phi_2 + \phi_3} \left ( R  + 2 (\partial \phi_1)^2 +  2 (\partial \phi_2)^2 + 2 (\partial \phi_3)^2 + 12  \right ) \label{ads5}
\ee
where ${\cal N} = {\cal N}_5  L_{y_1} L_{y_2} L_{y_3}$ with $L_{y_i}$ the length of the $y_i$ circles. The planar AdS$_5$ black hole
\be
ds^2 = \frac{dr^2}{r^2 \left ( 1 - \frac{m}{r^4} \right )} - r^2 \left ( 1 - \frac{m}{r^4} \right ) d t^2 + r^2 (dy_1^2 + dy_2^2 + dy_3^2),
\ee
then reduces to the asymptotically AdS$_2$/scalar solution given in the previous section, with $\alpha_1 = \alpha_2 = \alpha_3 = -1$. The thermodynamics is inherited from that of the 4d CFT, i.e. the onshell action scales as $T_H^3$. 

The action \eqref{ads5} admits a consistent truncation: one can set all the scalars to be equal
\be
\phi_1 = \phi_2 = \phi_3 = \phi
\ee
giving 
\be
 I = - {\cal N} \int d^2 x \sqrt{g} e^{ 3 \phi} \left ( R  + 6 (\partial \phi)^2 + 12  \right ). 
\ee
This truncation corresponds to solutions which respect the homogeneity of the compactification torus. 

\subsection{SYK and random averaging}

The geometries discussed in the previous section realise generalized conformal structure with specific values for the couplings of the scalar operators. A key feature of the SYK model
is of course that these couplings are drawn randomly from a Gaussian distribution and in this section we discuss the holographic implementation of this averaging. 

Formally we could implement the averaging as follows. From \eqref{asym-2} the scalar fields behave asymptotically as
\be
\Phi_A (\tau, \rho) = \rho^{1 - \Delta_A} \left ( \lambda_{A} (\tau) + \cdots \right ) 
\ee
and the non-normalizable modes $\lambda_A$ are the sources for the dual operators. Thus a bulk theory with boundary conditions $\lambda_A$ corresponds to a dual theory with fixed values for the operators sources. One could then implement the Gaussian averaging at the path integral level, by Gaussian averaging over bulk configurations with given boundary conditions. The problem with this approach is that it is far from clear what the IR behaviour of a theory with any given value of the couplings is, i.e. the interior behaviours of the individual bulk duals are not known. 

If we implement the Gaussian averaging directly in the SYK theory at large $N$, resulting in \eqref{eff2}, then the effective dilatation Ward identity can be expressed in terms of a 
single operator or, equivalently, in terms of equal sources for a large number of operators. For instance, in the $q=4$ case the dilatation Ward identity is
\be
{\cal H} = \frac{1}{4} {\cal O} \qquad \int dt {\cal O} (t) \equiv \sum_A \int dt {\cal O}_{A} (t) = \int dt_1 dt_2 \sum_{A} \tilde{\cal O}_A(t_1) \tilde{\cal O}_{A}(t_2)
\ee
where the operators $\tilde{\cal O}_A$ denote collectively ${\cal J} {\psi}_i \psi_j \psi_k \psi_l$.
 
To implement SYK holographically, we can therefore use two-dimensional gravity coupled to a single scalar or equivalently to a collection of scalars (with equal sources). The Ward identities are not however sufficient to infer the (leading) form for the bulk action. In higher dimensions, we could reconstruct the gravity/scalar interaction terms from knowledge of the two point functions but as we discussed earlier the correlation functions are trivial in this case. Thus we need other input to determine the gravity/scalar interactions - this comes from the solution of SYK, as the thermodynamic properties are consistent with underlying two-dimensional conformal invariance i.e. $\sigma  = 1$. 

This implies that both the Ward identities and the thermodynamics are consistent with a holographic action
\be
I = - {\cal N} \int d^{2} x \sqrt{g} \; e^{\sum_{A} \phi_A} \left ( R + 2  \right ) \label{mult-act2}
\ee 
with the following solution respecting the generalized conformal structure:
\be
ds^2 = \frac{1}{\rho^2} \left ( d \rho^2 - dt^2 \right ) \qquad 
e^{\phi_A} = \rho^{- \frac{1}{{\cal P}}}
\ee
where $ {\cal P}$ is the number of scalar fields/dual operators. 

For the purposes of discussing thermodynamics, and computing the Ward identities controlling the energy, we can effectively replace the multiple scalar fields by a single scalar field, as was done in previous literature. If however we include fields dual to other SYK operators, as well as those appearing in the Ward identities, as we do in section \ref{seven}, then we will need to take into account that they may couple differently to each of the $\phi_A$, and thus we cannot use just a single scalar field. 

\section{Reductions of CFTs to quantum mechanics with generalized conformal structure} \label{six}

\subsection{Toy example of CFT$_2$ reduction}

We begin by considering the toy example of a (classical) CFT$_2$ reduced on a circle, to illustrate the emergence of generalized conformal structure upon dimensional reduction. Consider a doublet of fermions $\psi^{\alpha}$ with Lagrangian
\be
I = - \frac{1}{2 \pi} \int d^2 x \sqrt{-h} \left ( \bar{\psi}^{\alpha} \rho^{i} D_i \psi^{\alpha} + \lambda \bar{\psi}^1 \bar{\psi}^2 \psi^1 \psi^2 \right )
\ee
Here we work in Lorentzian signature and $\rho^i$ represent the two-dimensional Dirac matrices, which satisfy
\be
\{ \rho^i, \rho^j \} = 2 \eta^{ij}
\ee
while  $\lambda$ is dimensionless. This is of course closely related to the well-known Gross-Neveu model  \cite{Gross:1974jv} ($N$ Dirac fermions with quartic interactions).
The classical stress energy tensor on a flat background can be expressed onshell as
\be 
{\cal T}_{ij} = \frac{1}{4} \left ( \bar{\psi}^{\alpha} \rho_i \partial_j \psi^{\alpha} + \bar{\psi}^{\alpha} \rho_j \partial_i \psi^{\alpha}  + 2 \lambda \bar{\psi}^1 \bar{\psi}^2 {\psi}^1 {\psi}^2 g_{ij} \right ), 
\ee
which is indeed traceless. At the quantum level, it is well-known that the theory is not conformal \cite{Gross:1974jv} but the classical theory can still be used as a useful warm-up example to illustrate how generalized conformal structure arises from the dimensional reduction.  

A convenient representation for the Dirac matrices is
\be
\rho^0 = \left( \begin{array}{cc}
0 & 1 \\
-1 & 0   \end{array} \right) \qquad
\rho^1 = \left( \begin{array}{cc}
0 & 1 \\
1 & 0   \end{array} \right)
\ee 
with the components of the Majorana fermions being 
\be
\psi^{\alpha} = \left( \begin{array}{c}
\psi^{\alpha}_- \\
\psi^{\alpha}_+   \end{array} \right) \
\ee 
The Lagrangian on a flat background can be expressed in terms of these components as 
\be
I = - \frac{1}{2 \pi} \int d^2 x  \left ( \psi_{+}^{\alpha} \partial_{+} \psi_{+}^{\alpha} + \psi_{-}^{\alpha} \partial_{-} \psi_{-}^{\alpha} 
+ \lambda {\psi}_+^1 {\psi}_+^2 \psi_-^1 \psi_-^2 \right ),
\ee
where $\partial_{\pm} = ( \partial_{t} \pm \partial_y )$.

Now let us consider the dimensional reduction of this model. We will need to distinguish between Neveu-Schwarz and Ramond sectors: in the former case
the fermion modes are integral while in the latter they are half integral. Let us write the harmonics of the fermions on the
circle as
\be
\psi^{\alpha}_{\pm} =  \sum_k \psi_{\pm}^{(k) \alpha} e^{\frac{i k y}{L}},
\ee
where the length of the $y$ circle is $2 \pi L$, and implicitly we impose a reality relation on $\psi_{\pm}^{\alpha}$.  

For integer moded expansions, the lowest modes are $k=0$ and retaining just these modes results in a reduced action:
\be
I = - L  \int d t   \left ( \psi_{+}^{(0) \alpha} \partial_{t} \psi_{+}^{(0)\alpha} + \psi_{-}^{(0) \alpha} \partial_{t} \psi_{-}^{(0) \alpha} 
+ \lambda {\psi}_+^{(0)1} {\psi}_+^{(0) 2} \psi_-^{(0)1} \psi_-^{(0)2} \right )
\ee
Rescaling the fermions as
\be
\chi_{\pm}^{\alpha} = \sqrt{L} \psi_{\pm}^{\alpha}
\ee
then gives
\be
I = -   \int d t   \left ( \chi_{+}^{ \alpha} \partial_{t} \chi_{+}^{\alpha} + \chi_{-}^{\alpha} \partial_{t} \chi_{-}^{ \alpha} 
+ \tilde{\lambda} {\chi}_+^{1} {\chi}_+^{ 2} \chi_-^{1} \chi_-^{2} \right ) \label{fer2}
\ee 
where $\chi$ is now dimensionless and $\tilde{\lambda}$ has mass dimension one. Reducing the stress energy tensor we then obtain 
\bea
{\cal O} &=&  \frac{1}{2}  ( \tilde{\lambda} {\chi}_+^{1} {\chi}_+^{ 2} \chi_-^{1} \chi_-^{2} ) \\
{\cal J} &=& \frac{1}{4} \left ( \chi_{+}^{\alpha} \partial_t \chi_{+}^{\alpha} - \chi_{-}^{\alpha} \partial_t \chi_{-}^{\alpha} \right ) \nn \\
{\cal H} &=& - \frac{1}{2}  \left ( \chi_{+}^{\alpha} \partial_t \chi_{+}^{\alpha} + \chi_{-}^{\alpha} \partial_t \chi_{-}^{\alpha} +  \tilde{\lambda} {\chi}_+^{1} {\chi}_+^{ 2} \chi_-^{1} \chi_-^{2}  \right ) \nn
\eea
Using the equations of motion e.g. 
\be
\partial_{t} \chi_{+}^1 + \frac{1}{2} \tilde{\lambda} \chi_{+}^2 \chi_{-}^1 \chi_{-}^2 = 0
\ee
the latter expression can be simplified onshell to give
\be
{\cal H} =  \frac{1}{2}  ( \tilde{\lambda} {\chi}_+^{1} {\chi}_+^{ 2} \chi_-^{1} \chi_-^{2} ) 
\ee
i.e. the Hamiltonian would be trivially zero if the coupling vanished. 

Thus the reduced theory automatically inherits from the trace and dilatation Ward identities of the classical two-dimensional CFT the relations
\be
{\cal H} = {\cal O} \qquad
\partial_{t} {\cal H} = \partial_{t} {\cal J} = 0. 
\ee
The reduced action \eqref{fer2} is clearly closely related to that of SYK - it is the SYK model with only four species of fermions and a specified quartic coupling. 
Just as for SYK, the theory could be solved perturbatively for small $\tilde{\lambda}$ by perturbing around free fields
(regulating the dimension) and perturbatively for large $\tilde{\lambda}$ by rescaling the fermions once again so that the kinetic term is treated perturbatively i.e. the action is
\be
I = -  \int d t   \left ( \frac{1}{\tilde{\lambda}^{\frac{1}{2} }} \tilde\chi_{+}^{ \alpha} \partial_{t} \tilde\chi_{+}^{\alpha} + \frac{1}{\tilde{\lambda}^{\frac{1}{2} }} \tilde\chi_{-}^{\alpha} \partial_{t} \tilde\chi_{-}^{ \alpha} 
+  {\tilde\chi}_+^{1} {\tilde\chi}_+^{ 2} \tilde\chi_-^{1} \tilde\chi_-^{2} \right )
\ee
and the path integral will be dominated by the interaction term in this limit.  

However, it only makes sense to neglect the Kaluza-Klein modes if one probes the behaviour at energies small compared to $1/L$ or, equivalently, probes distances
which are large compared to $1/\tilde{\lambda}$. Therefore, the KK modes are negligible for the dynamics on scales such that $t \tilde{\lambda} \gg 1$, in which limit the path integral is indeed dominated by the interaction term. 

\subsection{Reductions of higher-dimensional CFTs}

In this section we discuss the holographic dual of the AdS$_5$ reduced on a $T^3$, showing the emergence of generalized conformal structure in the one-dimensional reduced theory.  In this case the parent theory is conformal at the quantum level, and therefore the reduced theory will inherit generalized conformal structure at the quantum level.
The Euclidean ${\cal N} = 4$ SYM action is 
\be
I =  \int d^4 x \sqrt{g} \left (  \frac{1}{g_{YM}^2} {\rm Tr} (F^2 ) + {\rm Tr} (DX^a D X^a) - g_{YM}^2 {\rm Tr} [X^a,X^b]^2 + \cdots \right )
\ee
where $a =1,\cdots 6$. Here the ellipses denote the fermionic terms and we implicitly assume that the metric is flat, by not including curvature couplings. 
We now reduce on a torus of volume $L^3$ using the following ansatz for the metric
\be
ds^2 = ds_1^2 + e^{2 \Phi_1} dx_1^2 + e^{2 \Phi_2} dx_2^2 + e^{2 \Phi_3} dx_3^2 \equiv ds_1^2 + \sum_{i=1}^3 e^{2 \Phi_i} dx_i \cdot dx_i. 
\ee
Defining $X^m = \{ X^a, X^i \}$ where
\be
X^i = \frac{A_i}{\sqrt{2} g_{YM}}
\ee
the reduced action can be written as
\bea
I &=& L^3 \int d \tau \sqrt{g_1} e^{ \sum_i \Phi_i} \left (  {\rm Tr} (D_\tau X^a D^\tau X^a) + {\rm Tr} (D_\tau X^i D^{\tau} X^i) e^{-2 \Phi_i}   \right . 
\\ && \qquad  \left . - g_{YM}^2 \left ( {\rm Tr} [X^a,X^b]^2 + e^{-2 \Phi_i} {\rm Tr} [X^a,X^i]^2 + e^{-2 \Phi_i - 2 \Phi_j } {\rm Tr} [X^i,X^j]^2 \right ) + \cdots  \right ), \nn
\eea
where implicitly repeated indices $(a,b,i,j)$ are summed over. The equations of motion are
\bea
&& D_{\tau} D^{\tau} X^a + 2 g_{YM}^2 X^b [X^a,X^b]  + g_{YM}^2 e^{-2 \Phi_i} X^i [X^a,X^i] = 0; \\
&& D_{\tau} D^{\tau} X^i + g_{YM}^2 X^a [X^i,X^a]  + 2 g_{YM}^2 e^{- 2 \Phi_j} X^j [X^i,X^j] = 0. \nn
\eea
Writing the action as 
\be
I = L^3 \int d \tau \sqrt{g_1} {\cal L}
\ee
the operators dual to the couplings $\Phi_i$ are defined as
\bea
 {\cal O}_{\Phi_i}  &=& L^3 e^{2 \Phi_i - \sum_j \Phi_j} \frac{\delta {\cal L}}{\delta \Phi_i} \\
&=& L^3  e^{2 \Phi_i} \left (  - {\rm Tr} (X^a D_{\tau} D^\tau X^a) + {\rm Tr} (X^i D_{\tau} D^{\tau} X^i) e^{-2 \Phi_i}   
- \sum_{j \neq i} {\rm Tr} (X^j D_{\tau} D^{\tau} X^j) e^{-2 \Phi_j} \right . \nn \\ 
&& \qquad   - g_{YM}^2 \left ( {\rm Tr} [X^a,X^b]^2 - e^{-2 \Phi_i} {\rm Tr} [X^a,X^i]^2 + \sum_{j \neq i} e^{-2 \Phi_j} {\rm Tr} [X^a,X^j]^2   \right . \nn \\
&& \qquad \qquad \qquad  \left . \left .  -  \sum_{j \neq i } e^{-2 \Phi_i - 2 \Phi_j } {\rm Tr} [X^i,X^j]^2 +  \sum_{j,k \neq i } e^{-2 \Phi_j - 2 \Phi_k } {\rm Tr} [X^j,X^k]^2 \right )  \right ), \nn
\eea
where the summations over $a$ are implicit but the summations over $i$ are shown explicitly. Here we have included the boundary contributions to the action to make the onshell behaviour manifest. 

The Hamiltonian is 
\bea
{\cal H} &=& 2 L^3  e^{- \sum_j \Phi_j} \frac{1}{\sqrt{g_1}} \frac{\delta ( \sqrt{g_1}{\cal L})} {\delta g^{\tau \tau}}  \\
&=&    L^3 g_{\tau \tau}  \left ( {\rm Tr} (X^a D_{\tau} D^\tau X^a) + {\rm Tr} (X^i D_{\tau} D^{\tau} X^i) e^{-2 \Phi_i} + \right . \nn \\ 
&& \qquad  \left . - g_{YM}^2 \left ( {\rm Tr} [X^a,X^b]^2 + e^{-2 \Phi_i} {\rm Tr} [X^a,X^i]^2 
+  e^{-2 \Phi_i - 2 \Phi_j } {\rm Tr} [X^i,X^j]^2 \right ) + \cdots  \right ). \nn
\eea
Then the Ward identity
\be
g^{\tau \tau} {\cal H} + \sum_i e^{- 2 \Phi_i} {\cal O}_{\Phi_i} = 0  \label{ymward}
\ee
is satisfied onshell. This Ward identity \eqref{ymward} is the expected reduction of the trace relation for the 4d stress energy tensor, and is the field theory manifestation of the identity
found in the holographic model.  

When the couplings $\Phi_i$ are constant, they may be set to one (or, equivalently, absorbed into the definitions of $X^i$) so that the action can be written as 
\be
I = L^3 \int d \tau \sqrt{g_1}  \left (  {\rm Tr} (D_\tau X^m D^\tau X^m)  - g_{YM}^2 {\rm Tr} [X^m,X^n]^2  + \cdots  \right ). 
\ee
This is equivalent to the SYM action in one dimension, as it must be. However, D3-branes reduced on a $T^3$ and D0-branes differ in terms of what quantities are held 
fixed. For the former, we vary the 't Hooft coupling $g_{YM}^2 N$, while holding $N$ and $L$ fixed (in units of $\alpha'$), and the KK modes on $T^3$ are negligible provided we are only interested in energy scales $E \ll 1/L$. At finite temperature $T \ll 1/L$, the scalars $X^m \sim T$ and thus the entropy scales as $S \sim c (g_{YM}^2 N,N) L^3 T^3$ where the function $c$ depends on the 't Hooft coupling and the rank $N$.  

On the other hand, for the matrix model describing D0-branes in the decoupling limit, it is natural to write the action in terms of rescaled scalars as 
\be
I = \int d \tau \sqrt{g_1}  \left (  {\rm Tr} (D_\tau Y^m D^\tau Y^m)  - g_{0}^2 {\rm Tr} [Y^m,Y^n]^2  + \cdots  \right )
\ee 
where the coupling $g_{0}^2$  has (mass) dimension three and is related to the string coupling $g_s$ as 
\be
g_0^2 = \frac{g_s}{4 \pi^2 (\alpha')^{\frac{3}{2}}}. 
\ee
The appropriate limit is then $g_s \rightarrow 0$, $\alpha' \rightarrow 0$, $N \rightarrow \infty$  
with $g_0^2 N$ fixed. Since this coupling is dimensionful, we introduce an effective dimensionless coupling 
\be
\lambda (E) = \frac{g_0^2 N}{E^3} 
\ee
at a given energy scale $E$. When this effective 't Hooft coupling is small, perturbation theory is applicable, while in the opposite limit $\lambda(E) \gg 1$ the supergravity description in terms of decoupled D0-branes holds. Here the entropy scales as $S \sim c (\lambda(T),N)$ where again the function $c$ depends on the dimensionless coupling and the rank $N$. In this case, the scaling of the entropy with temperature is different at weak and strong coupling.

\section{Chaos in non-conformal theories} \label{seven}

\subsection{Two-point functions} \label{seven-one}

In this section we consider the holographic description of a scalar operator that is not driving the dimensionally driven flow. 
We first note that the couplings of bulk scalars to the background metric and running scalars are ambiguous without further input.  
We then show that these couplings are determined uniquely by the requirement that they respect the generalized conformal structure.
Such a theory can automatically be obtained by generalized toroidal 
reduction of a parent higher-dimensional AdS theory and the mass/dimension relation between bulk scalar/dual operator will be 
inherited from the parent AdS theory. 

In theories with a top-down realization such as non-conformal brane theories, the generalized conformal structure should be respected by all bulk excitations. 
It would be interesting to verify that bulk scalar excitations in non-conformal brane theories indeed do have couplings to the running scalar that are consistent with the generalized conformal structure and hence that there is a parent AdS theory. The cubic couplings in the bulk dual for SYK were previously discussed in \cite{Gross:2017hcz}. 
 
\bigskip

A natural proposal for describing a scalar operator with a two point function scaling as $\tau^{-2 \Delta_{\varphi}}$ would be a bulk action
\be
I _{\varphi} =  {\cal N} \int d^2 x \sqrt{g}  \;  \left ( (\partial \varphi)^2 + m_{\varphi}^2 \varphi^2 + \cdots \right ) \label{poss1}
\ee
coupling to the metric in dual frame, with ${\cal N}$ an appropriate normalization factor. Since the metric in the dual frame is asymptotically AdS$_2$, then, by the usual AdS$_2$/CFT$_1$ rules, the choice of $m_{\varphi}^2 = \Delta_{\varphi} (\Delta_{\varphi} - 1)$ guarantees that the two point function in the ground state exhibits scaling behaviour. 

However, this is not the only option to obtain a two point function with scaling behaviour and, as we will show below, the lack of coupling to the running scalar in \eqref{poss1} violates the generalized conformal structure. 

The discussion on AdS$_3$ reduction in section \ref{four} showed that 
an operator ${\cal O}_{\varphi}$ which has a two point function that scales as $\tau^{1 - 2 \Delta'_{\varphi}}$ can be described by a bulk scalar such that 
\be
I_{\varphi} =  {\cal N} \int d^2 x \sqrt{g} \;  e^{\sum_{A} \phi_A} \left ( (\partial \varphi)^2 + m_{\varphi}^2 \varphi^2 + \cdots \right ) \label{poss2}
\ee
where now $m_{\varphi}^2 = \Delta'_{\varphi} (\Delta'_{\varphi}  - 2)$ i.e. determined by the AdS$_3$ mass/dimension relation. 

More generally, let us consider the following action
\be
I_{\varphi} =  {\cal N}' \int d^2 x \sqrt{g} \;  e^{\sum_{A} \delta_A \phi_A} \left ( (\partial \varphi)^2 + m_{\varphi}^2 \varphi^2 + \cdots \right ), \label{poss3}
\ee
where the coefficients $\delta_A$ are a priori arbitrary. Then the equation of motion in the (Euclidean) AdS$_2$/dilaton geometry is
\be
\rho^2 \partial_{\rho}^2 \varphi + \sum_A \delta_A \alpha_A \rho \partial_{\rho} \varphi + \rho^2 \partial_{\tau}^2 \varphi = m_{\varphi}^2 \varphi.
\ee
This equation is identical to the equation of motion for a scalar in AdS$_{(D+1)}$ where 
\be
D = 1 - \sum_ {A} \alpha_A \delta_A.
\ee
The mass/dimension relation in $D+1$ is the standard one i.e. $m_{\varphi}^2 = \Delta' (\Delta' - D)$ and thus the solutions behave asymptotically as
\be
\varphi(\tau,\rho) = \rho^{D- \Delta'} \left ( \varphi_{D- \Delta'} (\tau) + \cdots \right ) + \rho^{\Delta'} \left ( \varphi_{\Delta'} (\tau) + \cdots \right )
\ee
Following the standard approach to holographic renormalization \cite{Skenderis2002}, one can thus show that the required counterterms for the scalar action are
\be
I_{\rm ct}  =  {\cal N}' \int dx \sqrt{h} \; e^{\sum_{A} \delta_A \phi_A} \left ( \frac{D-\Delta'}{2} \varphi^2 + \cdots \right )
\ee 
and that the scalar one point function can be written as 
\be
\langle {\cal O}_{\varphi} \rangle = - 2 {\cal N}' (2 \Delta' - D) \varphi_{\Delta'} + \cdots 
\ee
 where the ellipses in the latter denote terms analytic in the source $\varphi_{D - \Delta'}$. 
 
Solving the (Euclidean) field equation in frequency space gives the following solution that is regular in the interior
\be
\varphi (\omega, \rho) = \varphi_{D - \Delta'} (\omega) \rho^{\frac{D}{2}} K_{\Delta' - \frac{1}{2}}(\omega \rho)
\ee 
where $K_{\alpha}(z)$ is the modified Bessel function of order $\alpha$. From the asymptotic expansion of the Bessel function as $\rho \rightarrow 0$, we can then read off the behaviour of the two point function in frequency space:
\be
\langle {\cal O}_{\varphi} (\omega) {\cal O}_{\varphi} (-\omega) \rangle \propto \omega^{2 \Delta' - D} + \cdots,
\ee
where the ellipses denote contact terms that arise when $2 \Delta' - D$ is an even integer. 
 
The correlator in position space is then obtained by an inverse Fourier transform i.e. 
\be
\langle {\cal O}_{\varphi} (\tau) {\cal O}_{\varphi} (0) \rangle \propto \int d \omega e^{ i \omega \tau} \omega^{2 \Delta' - D} \sim \tau^{D -1 - 2 \Delta'} = \tau^{- \sum_A \alpha _A \delta_A - 2 \Delta'}
\ee 
which agrees with the particular result given earlier for AdS$_3$ reductions. 

\bigskip
 
Note that for high mass the two point function behaves as
\be
\langle {\cal O}_{\varphi} (\tau) {\cal O}_{\varphi} (0) \rangle \propto \tau^{-2 m_{\varphi}}
\ee
i.e. the couplings to the background running scalars give negligible contributions to the scaling. 
For high mass, the correlator is also well approximated by a geodesic in the asymptotically AdS$_2$ metric:
\be
\langle {\cal O}_{\varphi} (\tau) {\cal O}_{\varphi}(0) \rangle \propto \exp (- L(\tau))
\ee
where $L(\tau)$ is the renormalized length of a geodesic extending between the operators at the conformal boundary. The different possibilities for coupling to background scalars would only differ by subleading terms in the large mass limit. 

\bigskip

Following the discussions in section \ref{three}, we should note that the operator dimension cannot be read off immediately from the scaling in $\tau$, as the normalization of the supergravity 
action is itself dimensionful i.e. the coefficients ${\cal N}$ and ${\cal N}'$ appearing in \eqref{poss1}, \eqref{poss2} and \eqref{poss3} are dimensionful. In a holographic theory these normalisation constants take the form 
\be
{\cal N} = N^{2a} L^{2 b}
\ee
where $N \gg 1$ is related to the number of degrees of freedom in the dual theory, $L$ is a length scale (related to the dimensionful coupling of the dual theory) and $(a,b)$ are constants. (Loop and $\alpha'$ corrections to the supergravity action give contributions at lower orders in $N$ and the dimensionful 't Hooft coupling respectively.)
One can however always redefine the dual operator as ${\cal O}'_{\varphi}$, absorbing a factor of $L^{b}$, so that the two point function behaves as 
\be
\langle {\cal O}'_{\varphi} (\tau) {\cal O}'_{\varphi} (0) \rangle \sim N^{2a} \tau^{-2 m_{\varphi}}. \label{reddef}
\ee
Note that the higher derivative corrections to supergravity will contribute to this two point function as a power series in $L/\tau$, reflecting the underlying generalized conformal structure.

\bigskip

We now argue that generalized conformal structure uniquely fixes the couplings $\delta_A$ in \eqref{poss3}. This argument applies not just to the two-dimensional models under consideration here but more generally to any holographic model with generalized conformal structure and we will thus phrase our arguments for general dimensions. 

\FIGURE[t]{\includegraphics*{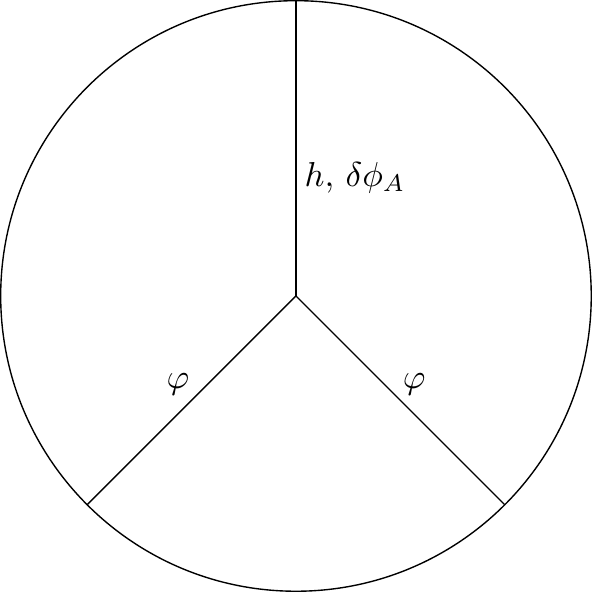}
\caption{Exchange diagrams for three point functions.}
\label{fig:three}}

We noted earlier that the trace Ward identity underlying the generalized conformal structure
\be
\langle {\cal T}^i_i \rangle + \sum_A \alpha_A \langle {\cal O}_{\phi_A} \rangle \sim 0,
\ee
where ${\cal T}^i_i$ is the trace of the stress energy tensor, 
implies an infinite tower of relations between correlation functions. In particular, it implies a relation between three point functions:
\be
\langle {\cal T}^i_i {\cal O}_{\varphi} {\cal O}_{\varphi} \rangle + \sum_A \alpha_A \langle {\cal O}_{\phi_A} {\cal O}_{\varphi} {\cal O}_{\varphi} \rangle \sim 0, \label{3pf2}
\ee
for a scalar operator ${\cal O}_{\varphi}$ dual to a bulk scalar field $\varphi$. However, these three point functions are computed holographically using the three point couplings in 
\be
I_{\varphi} = {\cal N}' \int d^{d+1} x \sqrt{g} \;  e^{\sum_{A} \delta_A \phi_A} \left ( (\partial \varphi)^2 + m_{\varphi}^2 \varphi^2 + \cdots \right ), \label{poss4}
\ee
between graviton/scalar and running scalars/scalar respectively. 

In particular, one perturbs the metric and the running scalars around their background values $\bar{g}$ and $\bar{\phi}_A$ as
\be
g = \bar{g} + h \qquad 
\phi_A = \bar{\phi}_A + \delta \phi_A
\ee
to obtain the cubic interactions between $(h,\varphi,\varphi)$ and $(\delta \phi_A, \varphi,\varphi)$. One then calculates the three point functions from the appropriate 
Witten diagrams shown in Figure~\ref{fig:three}, using the bulk to boundary propagators for $(h,\delta \phi_A, \varphi)$. 

\FIGURE[t]{\includegraphics*{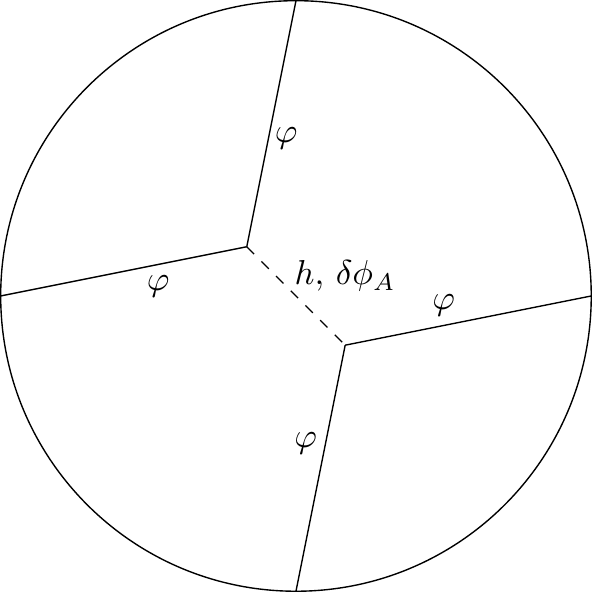}
\caption{Exchange diagrams for four point functions.}
\label{fig:two}}

The relation \eqref{3pf2} is reproduced only for the specific choice of the couplings $\delta_A =1$. In this case, the relation \eqref{3pf2} is automatically reproduced, as the 
action \eqref{poss4} can be obtained from the toroidal reduction of an action in $(2\sigma + 1)$ dimensions
\be
I_{\varphi} \sim  \int d^{2 \sigma +1} x \sqrt{G} \;  \left ( (\partial \varphi)^2 + m_{\varphi}^2 \varphi^2 + \cdots \right ), \label{poss5}
\ee
in which only the zero modes are retained. In the parent AdS theory, $\langle {\cal T}^a_a {\cal O}_{\varphi} {\cal O}_{\varphi} \rangle \sim 0$ and the reduction of this identity on the torus gives the required relation \eqref{3pf2}. 

In $d=1$ the three point functions in \eqref{3pf2} vanish (up to contact terms) but the three point couplings in \eqref{poss3} are needed to compute four point (and higher point) functions. Contributions to four point functions from Witten diagrams of the type shown in Figure ~\ref{fig:two} again respect the generalized conformal structure only for the specific choice of $\delta_A = 1$.

\subsection{Chaos and four-point functions in conformal theories} 
 
Let us begin by reviewing the relation of chaos to four-point functions in conformal field theories dual to AdS gravity, following \cite{Maldacena:2015waa}. Let $V$ and $W$ be generic operators in the dual field theory, such that neither acquire thermal expectation values. Now consider the following correlator 
\be
F(t) = {\rm Tr} \left ( y V(0) y W(t) y V(0) y W(t) \right )
\ee
where $y$ moves an operator a quarter of the way around the thermal circle i.e. 
\be
y^4 = \frac{1}{Z} e^{ - \beta H}. 
\ee
We define the disconnected correlator as 
\be
F_D(t) = {\rm Tr} \left ( y^2 V(0) y^2 V(0)  \right ) {\rm Tr} \left ( y^2  W(t) y^2 W(t) \right ),
\ee 
and the (dimensionless) function $f(t)$ as 
\be
f(t) = \frac{F(t)}{F_D(t)}. \label{chaos-def}
\ee
The underlying conformal symmetry implies that the function $f(t)$ is a function of the dimensionless ratio $t/\beta$ where $\beta = 1/T_H$. 

After the dissipation time $t_D \sim \beta$ but well before the scrambling time of the black hole, the out of time correlator $F(t)$ is approximated by the factorized value $F_D(t)$. 
For large times ($t \gg \beta$), CFTs holographically described by Einstein gravity give 
\be
f(t) = 1 - \epsilon \exp \left ( \frac{2 \pi t}{\beta}  \right ) + \cdots \label{chaos}
\ee
where $\epsilon$ is a small positive parameter proportional to the (dimensionless) Newton constant $G_{d+1}$ in AdS$_{d+1}$. gravity The exponential growth of the second term follows from the fact that the connected contributions to the correlator are dominated by high energy scattering near the horizon; the centre of mass energy for such processes grows exponentially. This term determines the Lyapunov exponent of the chaotic behaviour. 

One can understand the prefactor of the second term in \eqref{chaos} from the supergravity action:
\be
I = - \frac{1}{16 \pi G_{d+1}} \int d^{d+1} x \sqrt{g} \left  ( R + d (d+1) + \cdots \right )
\ee
The (connected) correlators computed from supergravity all have normalizations proportional to $1/G_{d+1}$.  For example, for AdS$_5$ for which $1/G_{5} \sim N^2$ all connected correlation functions scale as $N^2$. Thus for any four point function the connected contributions scale as $1/G_{d+1}$ while the disconnected contributions scale as the square of two point functions, i.e. as $1/G_{d+1}^2$. The second term in \eqref{chaos} follows from the ratio of a connected to disconnected contribution and therefore scales as $G_{d+1}$. 

Note that $G_{d+1}$ does not scale as $1/N^2$ in all dimensions (cf the claims in \cite{Maldacena:2015waa}). $G_7$ scales as $1/N^3$; $G_4$ scales as $1/N^{3/2}$ while $G_3$ scales as $1/N$. The latter scaling is in agreement with the scaling of the connected contribution in the SYK model.  
 
\subsection{Chaos and four point functions in non-conformal theories}

In this section we consider chaotic behaviour in a holographic theory with generalized conformal structure. Diffusion and chaos in NAdS$_2$ and related backgrounds has previously been discussed in \cite{Jensen:2016pah,Blake:2016jnn,Cotler:2016fpe,Davison:2016ngz} but here we emphasise the role of the underlying symmetry. 

We consider first the (Euclidean) action
\bea
I &=& - N^a L^b \int d^{d+1} x \sqrt{g} e^{\gamma \phi} \left ( R  + \beta (\partial \phi)^2 + C \right ) \label{nc-act1} \\
&& \qquad + N^a \int d^{d +1} x \sqrt{g} e^{\gamma \phi} \left (  (\partial \varphi_V)^2 + m_V^2 \varphi_V^2 + (\partial \varphi_W)^2 + m_W^2 \varphi_W^2 + \cdots \right ). \nn
\eea
As in \eqref{ingmar0} the metric and dilaton $\phi$ generate the non-conformal background while the scalar fields $\varphi_V$ and $\varphi_W$ are dual to operators $V$ and $W$ respectively. The ellipses include interaction terms for the scalar fields. 
In this section we work in general dimensions, but for simplicity consider a single scalar driving the non-conformal flow. The couplings of $\varphi_V$ and $\varphi_W$  
to this background scalar respect generalized conformal structure, as discussed in section \ref{seven-one}. 

The dilaton gravity part of the action is normalized with a dimensionful prefactor $L^b$ such that the scaling dimension of the operator dual to the graviton is $d$, as expected for a $d$-dimensional quantum field theory. For the other scalars, the dimensionful prefactor is absorbed into the definition of the scalar and the dual operator, as in \eqref{reddef}. The $N$ scaling is the same for all supergravity terms; terms arising from higher derivative corrections are subleading in $N$. 

The onset of chaos in a black hole background can then again be explored using the normalised four point function \eqref{chaos-def}. Compared to the conformal case, there are now three dimensionful scales in the problem $(L,t,\beta)$ and thus the Lyapunov exponent could in principle depend on all three. (The effective Newton constant depends explicitly on $L$.)
However, this action can always be expressed in terms of an action in $(2 \sigma + 1)$ dimensions, using the uplift \eqref{ingmar2}:
\bea
I &=& - c N^a \int d^{2 \sigma +1} x \sqrt{G}  \left ( R(G) + 2 \sigma (2 \sigma + 1) \right . \\
&& \qquad \qquad \qquad \qquad \qquad \left . -   (\partial \psi_V)^2 - m_V^2 \psi_V^2 - (\partial \psi_W)^2 - m_W^2 \psi_W^2 + \cdots \right ). \nn
\eea
Here $c$ is a numerical constant; $L^b$ is the volume of the $(2 \sigma - d)$-dimensional torus and $(\psi_V,\psi_W)$ are the uplifts of the scalar fields. The latter are rescaled by appropriate factors of $L$ relative to the lower-dimensional fields. 

In the uplifted theory the normalised four point function behaves as \eqref{chaos}, with $\epsilon \sim N^{-a}$. The dependence on the uplift dimension is implicit, via the temperature. The uplifted black hole background is 
\be
ds^2 = \frac{dr^2}{r^2 \left  (1 - \frac{m}{r^{2 \sigma}} \right )} - r^2 \left ( 1 - \frac{m}{r^{2 \sigma}}\right )dt^2 + r^2 dx \cdot dx_{(d-1)} + r^2 dz \cdot dz_{(2 \sigma -d)}
\ee
where the temperature is as given in \eqref{temp2} and depends on $\sigma$. The chaotic behaviour is associated with scattering within the $(t,r)$ plane close to the horizon; the only effect of the other directions is via the blackening function and hence the temperature. 

The action \eqref{nc-act1} assumes that the scalar fields $\varphi_V$ and $\varphi_W$ couple to the dilaton in such a way that a (formal) uplift to $(2 \sigma + 1)$ dimensions exists. It is not known whether this is indeed the case for generic fields in non-conformal brane theories, although one would expect the generalized conformal structure to be respected.  Note that the Kaluza-Klein reductions on spheres of non-conformal brane systems have not been explicitly computed so even the couplings to the running scalar for BPS operators dual to supergravity excitations are not known. 

However, one does not in fact need to know the detailed dilaton couplings with the scalars to argue that the normalized four point function behaves as \eqref{chaos} with $\epsilon \sim N^{-a}$: the argument relies only on the exchange near the horizon i.e. the uplift of the dilaton/metric sector to an AdS theory. 

\bigskip

We can also give an argument for \eqref{chaos} directly from the downstairs non-conformal structure as follows. The black hole metric in the dual frame is asymptotically AdS i.e.
\be
ds^2 = \frac{dr^2}{r^2 \left  (1 - \frac{m}{r^{2 \sigma}} \right )} - r^2 \left ( 1 - \frac{m}{r^{2 \sigma}}\right )dt^2 + r^2 dx \cdot dx_{(d-1)}.
\ee
The metric can be rewritten in Kruskal coordinates as 
\be
ds^2 = \left ( - A (uv) du dv + r^2 dx \cdot dx_{(d-1)} \right ) \qquad A(uv) = - \frac{4}{uv} \frac{ r^2 \left  (1 - \frac{m}{r^{2 \sigma}} \right )}{m^{\frac{1}{\sigma}} \sigma^2}
\ee
and the horizon is located at $u v = 0$. 

The growing term in the connected contribution to the correlator is associated with exchange near the $u=0$ horizon. Following the arguments of \cite{Shenker:2013pqa,Roberts:2014isa,Roberts:2014ifa,Maldacena:2015waa}, consider a perturbation of the fields near the boundary at large Killing time $t$. The energy associated with this perturbation behaves as $e L^b$ where $e$ is dimensionless and of order one. The dimensionful factor arises from the prefactor of the dilaton/gravity action: the black hole itself has energy that scales as $N^a L^b$, with the $N^a$ factor reflecting the scaling of the number of black hole microstates with the rank $N$ of the dual theory. 

Near the horizon the energy of the perturbation is boosted by a factor of $e^{2 \pi t/\beta}$. The boosted energy is thus comparable to the black hole energy when 
\be
t \sim \frac{\beta}{2 \pi} \log N^a
\ee
in agreement with the behaviour of the second term in \eqref{chaos}. So effectively the dimensional coupling $L^b$ controls the energy of both the black hole itself and of the propagating perturbation and this factor cancels out from the chaos time.

\section{Conclusions} \label{eight}

In this paper we have explored the role of generalized conformal structure in holographic duals to SYK and discussed more generally how generalized conformal structure determines interactions of bulk fields and chaotic behaviour in holographic models. 

We have shown that the underlying generalized conformal structure of SYK together with the thermodynamic properties indicates that the bulk description has a parent AdS$_3$ origin. This holds not just for the metric/running scalar sector but also for fields dual to generic operators used to probe chaos. In principle, one could bootstrap the bulk action by engineering the interactions to reproduce the leading contributions to all SYK correlation functions (see also \cite{Gross:2017hcz}). However, since the SYK model is a large $N$ model, this would result in an effective action involving a large number of fields (the number scaling as a power of $N$); this bulk action may have another interpretation in terms of, for example, a limit of a non-critical string theory. 

We have argued that in any holographic theory with generalized conformal structure, the couplings of bulk fields to the metric and running scalar are determined by the generalized conformal structure. It would be interesting to work out the supergravity spectrum and interactions in non-conformal brane theories, thus verifying that the generalized conformal structure is preserved by all the supergravity modes and their interactions. Holographic theories with generalized conformal structure can be interpreted in terms of generalized reductions of AdS theories in $(2 \sigma + 1)$ dimensions on $(2 \sigma -d)$ dimensional tori, where $\sigma$ is not necessarily integral. These parent AdS theories would essentially determine the dynamics of non-conformal brane theories at large 't Hooft coupling.  

We have constructed two-dimensional gravity theories with multiple scalars and shown that these also admit parent AdS theories, where the dimension of the parent theory is determined by the thermodynamic properties. These bulk theories include holographic duals to conformal theories compactified on spatial tori, such as the case of ${\cal N} = 4$ SYM compactified on a three torus discussed here. It would be interesting to explore applications of this general class of two dimensional models. Note that various dualities involving 2d bulk theories have recently been discussed in \cite{Mezei:2017kmw}. 

Our discussions of chaotic behaviour in holographic theories with generalized conformal structure are based on the arguments of \cite{Shenker:2013pqa,Roberts:2014isa,Roberts:2014ifa,Maldacena:2015waa} about exchange near the black hole horizon. It would be interesting to calculate the relevant four point function contributions explicitly holographically, by computing the Witten diagrams shown in Figure~\ref{fig:two}, using real-time holography methods \cite{Skenderis2008,Skenderis2009}.

\section*{Acknowledgments}

This work was supported by the Science and Technology Facilities Council (Consolidated Grant ``Exploring the Limits of the Standard Model and Beyond'').
We thank the Simons Center, the GGI and Kavli IPMU for hospitality during the completion of this work. This project has received funding from the European Union's Horizon 2020
research and innovation programme under the Marie Sklodowska-Curie grant
agreement No 690575.

\bibliographystyle{jhep2}

\providecommand{\href}[2]{#2}\begingroup\raggedright\endgroup

\end{document}